\documentclass[10pt,letterpaper,optengjnl]{article}

\usepackage{opteng}
\usepackage{times}
\usepackage[scaled=0.92]{helvet}
\usepackage[modulo,switch]{lineno}
\usepackage[
    pdftitle={GREGOR Fabry-Perot Interferometer - status report and prospects},
    pdfauthor={Klaus G. Puschmann, Horst Balthasar, Rohan E. Louis, Emil Popow,
        Manfred Woche, Carsten Denker, Christian Beck, Thomas Seelemann,
        and Reiner Volkmer},
    pdfsubject={Solar Physics},
    pdfkeywords={test, test, test},
    colorlinks=true,
    breaklinks=true,
    anchorcolor=blue,
    filecolor=blue,
    menucolor=blue,
    pagecolor=blue,
    linkcolor=blue,
    citecolor=blue,
    urlcolor=blue,
    final]{hyperref}

\fancyput(17.5cm,2.5cm){  \rput[l](-18.6,-1.25){\footnotesize \textsf{Optical Engineering 52(8),121432SSPR (\today)}}  \rput(-17.3,-27){\footnotesize \textsf{Optical Engineering}}  \rput(-2,-27){\footnotesize \textsf{\today/Vol.~52(8)}}}%
\definecolor{myRed}{rgb}{0.84,0.08,0.52}

\newcommand\arcsec{\mbox{$^{\prime\prime}$}}
\newcommand\phn{\phantom{0}}

\begin{document}
\twocolumn[

\title{The GREGOR Fabry-P\'erot Interferometer and its companion the Blue Imaging Solar Spectrometer\\
    }

\begin{minipage}{7cm} 

\author{Klaus G.\ Puschmann\\
    Carsten Denker\\
    Horst Balthasar\\
    Rohan E.\ Louis\\
    Emil Popow\\
    Manfred Woche}
    
\address{Leibniz-Institut f\"ur Astrophysik Potsdam\\
    14482 Potsdam\\
    Germany}

\email{\href{mailto:kgp@aip.de}{kgp@aip.de}}

\author{Christian Beck}

\address{Instituto de Astrof\'isica de Canarias\\
    38205 La Laguna, Tenerife\\
    Spain}

\author{Thomas Seelemann}

\address{LaVision\\
    37081 G\"ottingen\\
    Germany}

\author{Reiner Volkmer}

\address{Kiepenheuer-Institut f\"ur Sonnenphysik\\
    79104 Freiburg\\
    Germany}


\end{minipage}
\hfill%
\begin{minipage}{9cm}


%
%

\begin{abstract}
The GREGOR Fabry-P\'erot Interferometer (GFPI) is one of three first-light
instruments of the German 1.5-meter GREGOR solar telescope at the Observatorio
del Teide, Tenerife, Spain. The GFPI allows fast narrow-band imaging and
post-factum image restoration. The retrieved physical parameters will be a
fundamental building block for understanding the dynamic Sun and its magnetic
field at spatial scales down to $\sim 50$~km on the solar surface. The GFPI is a
tunable dual-etalon system in a collimated mounting. It is designed for
spectrometric and spectropolarimetric observations between 530--860~nm and 580--660~nm, 
respectively, and possesses a theoretical spectral 
resolution of ${\cal R}\approx 250,000$. Large-format, high-cadence CCD
detectors with sophisticated computer hard- and software enable the scanning of
spectral lines in time-spans equivalent to the evolution time of solar features.
The field-of-view (FOV) of $50\arcsec \times 38\arcsec$ covers a significant fraction
of the typical area of active regions in the spectroscopic mode. In case of Stokes-vector 
spectropolarimetry, the FOV reduces to $25\arcsec \times 38\arcsec$. We present the main 
characteristics of the GFPI including advanced and automated calibration and observing procedures.
We discuss improvements in the optical design of the instrument and show first
observational results. Finally, we lay out first concrete ideas for the
integration of a second FPI, the Blue Imaging Solar Spectrometer (BLISS), which will
explore the blue spectral region below 530~nm.
\end{abstract}


\end{minipage}
\vskip 1cm

]


%
%

\section{Introduction}\label{SEC01}

Solar physics has made tremendous progress during recent years thanks to
numerical simulations and high-resolution spectropolarimetric observations with
modern solar telescopes such as the Swedish Solar
Telescope\cite{2003SPIE.4853..341S}, the Solar Optical Telescope on board the
Japanese HINODE satellite,\cite{2007SoPh..243....3K} and the stratospheric
Sunrise telescope.\cite{2010ApJ...723L.127S} Taking the nature of sunspots as an
example, many important new observational results have been found, e.g., details
about the brightness of penumbral filaments, the Evershed flow, the dark-cored
penumbral filaments, the net circular polarization, and the moving magnetic
features in the sunspot moat. Telescopes with apertures of about 1.5~m such as
the GREGOR solar telescope\cite{2012ASPC..463..365S, 2012AN....333..796S} or the New Solar
Telescope\cite{2006SPIE.6267E..10D, 2010SPIE.7733E..93C} will help to
discriminate among competing sunspot models and to explain the energy balance of
sunspots. New results on the emergence, evolution, and disappearance of magnetic
flux at smallest scales can also be expected. However, these 1.5-meter
telescopes are just the precursors of the next-generation solar telescopes,
i.e., the Advanced Technology Solar Telescope\cite{2008SPIE.7012E..16W} and the
European Solar Telescope,\cite{2010SPIE.7733E..15C} which will finally be able
to resolve the fundamental scales of the solar photosphere, namely, the photon
mean free path and the pressure scale height.

Fabry-P\'erot interferometers (FPIs) have gained importance in solar
physics during the last decades because they deliver high spatial and spectral
resolution. A growing number of such instruments is in operation at various
telescopes. Although most of these instruments have been initially designed only
for spectroscopy, most of them have now been upgraded to provide full-Stokes
polarimetry.\cite{2011A&A...533A..21P, 2011SoPh..268...57M, 2010SPIE.7733E..14R}
The Universit\"ats-Sternwarte G\"ottingen developed an imaging spectrometer for
the German Vacuum Tower Telescope (VTT) in the early 1990s \cite{1992A&A...257..817B}. 

A fundamental renewal of the G\"ottingen FPI during the first half of 2005 was
the starting point of the development of a new FPI for the 1.5-meter GREGOR
telescope.\cite{2006A&A...451.1151P} New narrow-band etalons and new
large-format, high-cadence CCD detectors were integrated into the instrument,
accompanied by powerful computer hard- and software. From 2006 to 2007, the
optical design for the GREGOR Fabry-P\'erot Interferometer (GFPI) was developed,
the necessary optical elements were purchased, and the opto-mechanical mounts
were manufactured.\cite{2007msfa.conf...45P} An upgrade to full-Stokes
spectropolarimetry followed in 2008.\cite{2008A&A...480..265B,
2009IAUS..259..665B} In 2009, the Leibniz-Institut f\"ur
Astrophysik Potsdam took over the scientific responsibility for the GFPI, and
the instrument was finally installed at the GREGOR 
telescope.\cite{2010SPIE.7735E.217D} During the commissioning phase in 2011,
the software was prepared for TCP/IP communication with external devices according to the Device Communication Protocol
(DCP\cite{2012AN....333..840H}) and three computer-controlled translation stages (two filter sliders and one mirror
stage) were integrated into the GFPI.\cite{2012ASPC..463..423P} This permits automated observing and
calibration procedures and facilitates easy operations during observing runs.
The present article is a modified version of a contribution\cite{2012SPIE.8446E..79P} to the 
SPIE conference `Ground-Based and Airborne Instrumentation for Astronomy IV', which took place 
in Amsterdam in June 2012. A complementary description of the GFPI\cite{2012AN....333..880P} 
has been published in the meanwhile in a special issue of Astronomische Nachrichten/AN (Volume 333, Issue 9) dedicated to the 
GREGOR solar telescope.

%
%

\section{GREGOR solar telescope}\label{SEC02}

The 1.5-meter GREGOR telescope is the largest European solar telescope and is
designed for high-precision measurements of dynamic photospheric and
chromospheric structures and their magnetic field. Some key scientific topics of
the GREGOR telescope are: 
the interaction between convection and magnetic fields in the photosphere, and the 
enigmatic heating mechanism of the chromosphere. The inclusion of a
spectrograph for stellar activity studies and the search for solar twins expand
the scientific usage of the GREGOR telescope to the nighttime
domain.\cite{2012AN....333..901S}

The GREGOR telescope replaced the 45-cm Gregory-Coud\'e Telescope, which had
been operated on Tenerife since 1985. The construction of the new telescope was
carried out by a consortium of several German institutes, namely, the
Kiepenheuer-Institut f\"ur Sonnenphysik Freiburg, the Leibniz-Institut f\"ur
Astrophysik Potsdam (AIP), and the Institut f\"ur Astrophysik G\"ottingen. In
2009, the Max-Planck-Institut f\"ur Sonnensystemforschung in Katlenburg-Lindau
took over the contingent of the latter institute. The consortium maintains 
partnerships with the Instituto de Astrof\'isica de Canarias in Spain and the
Astronomical Institute of the Academy of Sciences of the Czech Republic in
Ond\v{r}ejov.

The GREGOR telescope is an alt-azimuthally mounted telescope with an open
structure and an actively cooled light-weighted Zerodur primary mirror. The
completely retractable dome\cite{2012AN....333..830H} enables wind 
flushing through the telescope to facilitate cooling of telescope structure and 
optics.\cite{2012AN....333..847S, 2012AN....333..816V}
The water-cooled field stop at the primary focus provides a field-of-view (FOV)
with a diameter of 150$^{\prime\prime}$. The light is reflected via two
elliptical mirrors and several flat mirrors into the optical laboratory. Passing the 
GREGOR adaptive optics system (GAOS),\cite{2012AN....333..863B} the light is finally distributed to the
scientific instruments. The removable GREGOR Polarimetric Unit 
(GPU)\cite{2012AN....333..854H}, which was developed at the AIP, is located 
near the secondary focus and ensures high-precision polarimetric 
calibration. In addition, a telescope polarization model has been developed,\cite{2011ASPC..437..351B} 
since the instrumental polarization is time-dependent due to the alt-azimuthal mount.
Three first-light instruments have been commissioned in 2011 and passed through science 
verification in 2012: the GRating
Infrared Spectrograph (GRIS),\cite{2012AN....333..872C} the Broad-Band Imager
(BBI),\cite{2012AN....333..894V} and the GFPI\cite{2012AN....333..880P}. 
GRIS and GFPI can be used simultaneously. A dichroic beamsplitter directs wavelengths above
650~nm to the spectrograph, whereas all shorter wavelengths are reflected
towards to the GFPI. In the near future, this beamsplitter can be exchanged with a different
one with a cutoff above 860~nm.

%
%

\begin{figure*}[!ht]
\centerline{\includegraphics[width=\textwidth]{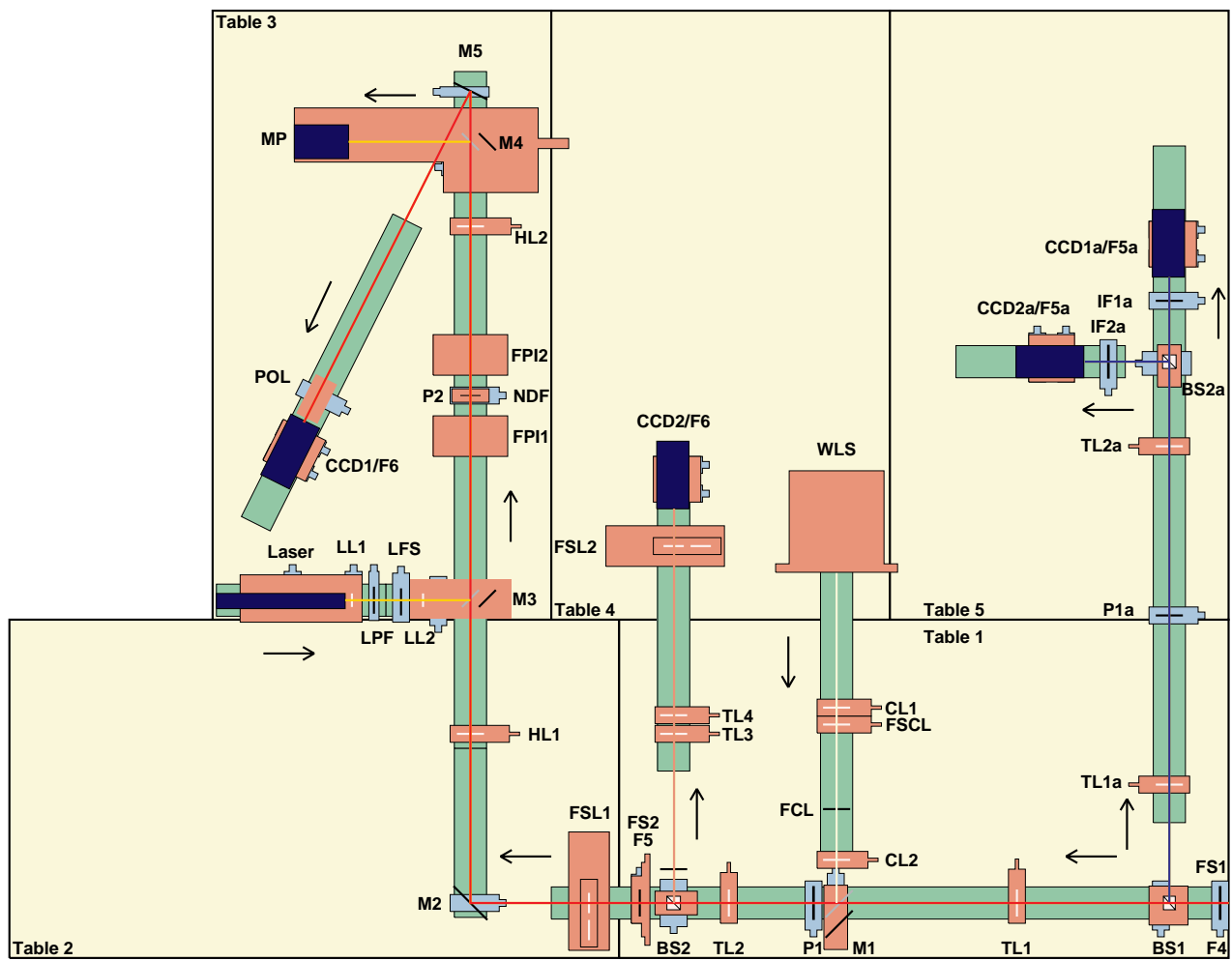}}
\caption{Up-to-scale drawing of the GFPI, including narrow-band channel (\textsf{NBC}, marked as red beam), 
    broad-band channel (\textsf{BBC}, salmon beam), blue imaging channel (\textsf{BIC}, blue beam), and all auxiliary channels, i.e., white-light channel (off-wight beam) and laser/photo-multiplier channel (yellow beam). The light from the telescope enters the instrument at the science focus \textsf{F4}. Black arrows denote the light direction in the individual channels. The optical elements of \textsf{BIC} are labeled with an extra ``a''. 
    \textsf{CCD1}, \textsf{CCD2}: Imager QE detectors;
    \textsf{CCD1a}, \textsf{CCD2a}: pco.4000 detectors;
    \textsf{FPI1}, \textsf{FPI2}: narrow-band etalons;
    \textsf{NDF}: neutral density filter;
    \textsf{TL1} ($f = 600$~mm, $d = 63$~mm),
        \textsf{TL2} ($f = 250$~mm, $d = 40$~mm),
        \textsf{TL1a} ($f = 500$~mm, $d = 63$~mm),
        \textsf{TL2a} ($f = 500$~mm, $d = 80$~mm),
        \textsf{TL3} ($f = 400$~mm, $d = 63$~mm),
        \textsf{TL4} ($f = 600$~mm, $d = 63$~mm),
        \textsf{HL1} ($f = 1000$~mm, $d = 80$~mm),
        \textsf{HL2} ($f = 1500$~mm, $d = 100$~mm): achromatic lenses;
    \textsf{CL1} ($f = 300$~mm, $d = 63$~mm),
        \textsf{CL2} ($f = 150$~mm, $d = 40$~mm): plano-convex lenses;
    \textsf{M1}, \textsf{M3}, \textsf{M4}: removable folding mirrors;
    \textsf{M2}, \textsf{M5}: fixed folding mirrors (60~mm $\times$ 85~mm);
    \textsf{F4}, \textsf{F5}, \textsf{F6}, \textsf{F5a},\textsf{FCL}: foci;
    \textsf{P1}, \textsf{P1a}, \textsf{P2}: pupil images;
    \textsf{BS1}, \textsf{BS2}, \textsf{BS2a}: beamsplitters (40~mm $\times$
        40~mm);
    \textsf{FS1}, \textsf{FS2}, \textsf{FSCL}: field stops;
    \textsf{WLS}: white-light source (slide projector);
    \textsf{FSL1}, \textsf{FSL2}: filter sliders;
    \textsf{IF1a}, \textsf{IF2a}: interference filters;
    \textsf{POL}: full-Stokes polarimeter;
    \textsf{LL1}, \textsf{LL2}: laser lenses;
    \textsf{LPF}: laser polarization filter;
    \textsf{LFS}: laser field stop; and
    \textsf{MP}: photomultiplier.}
\label{FIG01}
\end{figure*}

\section{GREGOR Fabry-Perot Interferometer}\label{SEC03}

\subsection{Optical design}\label{SEC03_1}

The GFPI is mounted on five optical tables and is protected by an aluminum
housing to prevent pollution of the optics by dust and to reduce
stray-light. The optical layout of the instrument is shown in Fig.~\ref{FIG01}.
Behind the science focus \textsf{F4}, four achromatic lenses \textsf{TL1},
\textsf{TL2}, \textsf{HL1}, \& \textsf{HL2} create two more foci \textsf{F5} \&
\textsf{F6} and two pupil images \textsf{P1} \& \textsf{P2} in the narrow-band
channel \textsf{NBC}. The etalons \textsf{FPI1} \& \textsf{FPI2} are placed in
the vicinity of the secondary pupil in the collimated light beam. A neutral
density filter \textsf{NDF} between the two etalons with a transmission of 63\%
reduces the inter-etalon reflexes. The beam in the \textsf{NBC} is folded twice
by \textsf{M2} \& \textsf{M5} at a distance of 500~mm and 400~mm from
\textsf{F5} and \textsf{HL2}, respectively, to minimize the instrument envelope.
Two field stops \textsf{FS1} \& \textsf{FS2} prevent an over-illumination of the detectors, 
where the secondary adjustable field stop especially avoids an overlap of the two images created by
the removable dual-beam full-Stokes polarimeter. A beamsplitter cube
\textsf{BS2} near \textsf{F5} directs 5\% of the light to the broad-band channel
\textsf{BBC}. There, the two achromatic lenses \textsf{TL3} \& \textsf{TL4} are
chosen such that the image scale of the detectors \textsf{CCD1} \& \textsf{CCD2}
is exactly the same. The simultaneous recording of broad- and narrow-band
images is a prerequisite for post-factum image reconstruction.
Several image reconstruction techniques are already implemented in a data pipeline 
presently being developed at the AIP.\cite{2012SPIE.8446E..79P}
A detailed description of these techniques and a comparison of related results 
has been recently published.\cite{2011A&A...533A..21P}
A dichroic beamsplitter \textsf{BS1} just behind the science focus \textsf{F4} 
sends the blue part of the spectrum (below 530~nm) to the Blue Imaging Channel \textsf{BIC}. 
One-to-one imaging with the lenses \textsf{TL1a} \& \textsf{TL2a} provides the option of recording 
broad-band images in the wavelength range 380-530~nm with two pco.4000 cameras \textsf{CCD1a} \&
\textsf{CCD2a} behind beamsplitter \textsf{BS2a} in parallel to \textsf{NBC} and \textsf{BBC} or 
GRIS observations. This imaging channel will be replaced in the future by the Blue Imaging Solar Spectrometer 
(BLISS), which is described in Section~\ref{BLISS}.

Three computer-controlled precision translation stages facilitate automated
observing sequences. The two stages \textsf{FSL1} \& \textsf{FSL2} are used to
switch between two sets of interference filters. The filters restrict the
bandpass of \textsf{BBC} and \textsf{NBC} to a full-width at half-maximum
(FWHM) of 10~nm and 0.3--0.8~nm, respectively. The near-focus position of the 
sliders excluded the usage of a filterwheel because of the required positioning 
accuracy in the order of micrometers. Only two different interference filters can be 
mounted on each slider because of the limited space on the optical tables. 
The pre-filters of each channel can be tilted to optimize the wavelength 
of the transmission maximum. At present, the optical elements of the GFPI 
are not super-achromatic. Thus, a chromatic focus shift restricts the maximum spectral distance between
two different pre-filters/spectral lines to about 100~nm\cite{2012AN....333..880P}. 
A third stage inserts a deflection mirror \textsf{M1} into the light path to take 
calibration data with a continuum-light source for spectral calibration purposes (white-light channel). 
A laser/photo-multiplier channel for finesse adjustment of the etalons completes the optical setup.

\subsection{Cameras, etalons, and control software}

\begin{table*}[t]
\begin{center}
\caption{Summary of the GFPI observations.}
\label{TAB01}
\small
\medskip
\begin{tabular}{lcccccccccc}
\hline\hline
Channel & $\lambda_0$ [nm] & $\Delta t$ [ms] & $I$ [counts] & \textsf{PHA}
    (\textsf{F4}) & \textsf{PH} (\textsf{F3}) & \textsf{PH} (\textsf{F2}) &
    \textsf{TG} & \textsf{QS} & \textsf{SP} & \textsf{PF}\rule[-2mm]{0mm}{6mm}\\
\hline
NBC ($1 \times 1$ binning) & 543.3 &     40 & \phn 2400 &  x &  x & -- & x & x &  x &  x\rule{0mm}{4mm}\\
BBC ($1 \times 1$ binning) & 543.3 &     40 &        -- &  x &  x & -- & x & x &  x & --\\
NBC ($2 \times 2$ binning) & 543.3 &     30 & \phn 3500 & -- & -- & -- & x & x & -- &  x\\
NBC ($2 \times 2$ binning) & 557.6 &     40 & \phn 2100 & -- &  x & -- & x & x & -- &  x\\
NBC ($2 \times 2$ binning) & 617.3 &     10 & \phn 3200 & -- &  x & -- & x & x & -- &  x\rule[-2mm]{0mm}{4mm}\\
\hline
Ca\,\textsc{ii}\,H & 396.8 &     15 & \phn 9230 &  x &  x &  x & x & x &  x & --\rule{0mm}{4mm}\\
G-band             & 430.7 & \phn 6 &     10350 &  x &  x &  x & x & x &  x & --\\
Blue continuum     & 450.6 & \phn 3 &     11300 &  x &  x &  x & x & x &  x & --\rule[-2mm]{0mm}{4mm}\\
\hline\vspace*{-2mm}
\end{tabular}
\parbox{0.87\textwidth}{Note. --- Central wavelength $\lambda_0$,
    exposure time $\Delta t$, and mean intensity $I$ for all observations
    (\textsf{PHA}: pinhole array, \textsf{PH}: pinhole, \textsf{TG}: target,
    \textsf{QS}: quiet Sun, \textsf{SP}: sunspot, and \textsf{PF}:
    pre-filter).}
\end{center}
\end{table*}

The GFPI data acquisition system consists of two Imager QE CCD cameras with Sony
ICX285AL detectors, which have a full-well capacity of 18,000~e$^{-}$ and a
read-out noise of 4.5~e$^{-}$. The detectors have a spectral response from
320--900~nm with a maximum quantum efficiency of $\sim$60\% at 550~nm. The chips
have $1376 \times 1040$ pixels with a size of 6.45~$\mu$m $\times$ 6.45~$\mu$m.
The image scale at both cameras is 0.0361\arcsec\ pixel$^{-1}$, which leads to a FOV of $50\arcsec \times
38\arcsec$ in the spectroscopic mode. Due to the dual-beam configuration, the 
FOV shrinks to $25\arcsec \times 38\arcsec$ when using the polarimetric 
mode of the instrument. The maximal blueshift due to the collimated mounting of the 
etalons is about 4.3~pm at 630~nm. The cameras are triggered by analog TTL signals of a 
programmable timing unit (PTU). The analog-digital conversion of the CCD-images is carried out with 
12-bit resolution. The data recorded by the cameras are passed via digital 
coaxial cables to the GFPI control computer and are stored on a RAID~0 system.\cite{2012ASPC..463..423P}

Two pco.4000 cameras in stock at the observatory can be used for imaging in the \textsf{BIC}. The pco.4000 camera has a full-well capacity of
60,000~e$^{-}$ and a read-out noise of 11~e$^{-}$. The detector has a spectral
response from 320--900~nm with a quantum efficiency of $\sim$32\% ($\sim$45\%)
at 380 (530)~nm. The chip has $4008\times 2672$ pixels, each with a size of 
9~$\mu$m $\times$ 9~$\mu$m. The image scale at both cameras is 0.031\arcsec\ pixel$^{-1}$. 
Because of a vignetting of the beam by \textsf{BS1} and \textsf{BS2a} only $2000 \times 2672$ pixels can be
used resulting in a FOV of $63\arcsec \times 84\arcsec$. Three interference
filters for Ca\,\textsc{ii}\,H $\lambda 396.8$~nm, Fraunhofer G-band $\lambda
430.7$~nm, and blue continuum $\lambda 450.6$~nm are available, which have a $\mathrm{FWHM} = 1$~nm
and a transmission better than 60\%.

The two GFPI etalons manufactured by IC Optical Systems (ICOS) have a diameter
$\oslash = 70$~mm, a measured finesse ${\cal F} \sim 46$, spacings $d = 1.1$
and 1.4~mm, and a high-reflectivity coating ($R \sim 95$\%) in the wavelength
range from 530--860~nm. The resulting FWHM of the instrument is
on the order of 1.9--5.6~pm and leads to a theoretical
spectral resolution of ${\cal R} \sim 250,000$. However, the etalon spacings are not optimal
because of historical reasons, thus off-band contributions
can become critical when using too broad pre-filters.\cite{2012AN....333..880P}
All etalons are operated by three-channel CS100 controllers manufactured by ICOS. The cavity spacings are
digitally controlled by the GFPI control computer via RS-232 communication. A
thermally insulated box protects pupil and etalons from stray-light and
air flows inside the instrument.

The communication between internal (cameras, etalons, and filter and mirror
sliders) and peripheral devices (telescope, AO system, AO filter wheel, GPU,
GRIS, etc.) is controlled by the software package DaVis from LaVision in
G\"ottingen, which has been adapted to the needs of the
spectrometer.\cite{2006A&A...451.1151P, 2012ASPC..463..423P} The modification of
the software for TCP/IP communication with external devices using DCP allows an
easy implementation of automated observing procedures.\cite{2012ASPC..463..423P} 
All observing modes such as etalon adjustment, line finding, flat-fielding, recording of dark, pinhole,
and target images, continuum scans, and recording of scientific data are now
automated.

\subsection{Polarimetry with the GFPI}
\label{polarimetry}
The science verification time in 2012 was mostly devoted to spectroscopy. 
Nevertheless, the GFPI is equipped with a dual-beam full-Stokes polarimeter\cite{2008A&A...480..265B, 2012AN....333..880P} 
that can be inserted in front of the detector in the \textsf{NBC}. The polarimeter consists of two 
ferro-electric liquid crystal retarders (FLCRs) and a modified Savart-plate. The first liquid crystal acts as a half-wave plate 
and the second one as a quarter-wave plate at a nominal wavelength of 630~nm. The modified Savart-plate consists
of two polarizing beamsplitters and an additional half-wave plate, which exchanges the ordinary and 
the extraordinary beam. With this configuration, the separation of the 
two beams is optimized and the orientation of the astigmatism in both beams is the same so that it 
can be corrected by a cylindrical lens. For the present set of FLCRs their properties (retardation and fast 
axis orientation) have been measured so far only for the spectral range 580--660~nm.\cite{2012AN....333..880P} An automated polarimetric calibration procedure will be integrated during an upcoming technical campaign in April 2013. At that time the GREGOR polarimetric calibration unit will be available for the visible-light range. This will be the starting point for polarimetry with the GFPI at the GREGOR solar telescope. 
A future enlargement of the reduced FOV in the polarimetric dual-beam mode would be possible by replacing the Imager QE cameras with modern and large-format sCMOS cameras, which have also much higher frame rates (cf. Sect.~\ref{BLISS}). However, this would require an additional re-design of the polarimeter, because presently the FOV is also limited by the size of the calcites of the polarimeter.

%
%

\section{GFPI science verification}\label{SEC04}

We took several data sets in a technical campaign from 15~May to 1~June 2012 
for a first characterization of the GFPI performance. The AO system was not
available because of technical problems with the control computer. Thus, our
efforts have been restricted to an optimization of the system and to
observations of test data as measures for intensity levels, image quality,
spectral resolution, stray-light, and other performance indicators of \textsf{NBC}, 
\textsf{BBC}, and \textsf{BIC}. Several problems related to the RS-232
communication with the FPI controllers were resolved so that a stable finesse is
now achieved over several days. In addition, the timing between cameras and
etalons has been optimized to ensure that images are only taken when the etalon
spacing has settled to its nominal value.

\subsection{Imaging spectrometric data}

\begin{table*}[t]
\begin{center}
\caption{NBC pre-filter characteristics.}
\label{TAB02}
\medskip
\small
{
\begin{tabular}{lcccccccc}
\hline\hline
Filter\rule[-2mm]{0mm}{6mm} & $\lambda_0$ [nm] & FWHM [nm] & $T$ [\%] & Binning &
   $\Delta t$ [ms] & $I$ [counts] & Frame rate [Hz]\\
\hline
ANDV11436       & 543.4 & 0.4 & 38 & - / - /  $2 \times 2$ & 60 / 100 / 30        & 1200 / 2000 / 3500           & \phn 7 / 5 / 11\rule{0mm}{4mm}\\
1100 BARR9      & 543.4 & 0.6 & 70 & - & 40 / 60     & 1700 / 2700           & \phn 7 / 6 \\
ANDV5288        & 557.6 & 0.3 & 40 & $2 \times 2$ & 40 & 2100 &     11 \\
DV5289 AM-32389 & 569.1 & 0.3 & 45 & - & 60 / 100         & 1200 / 2000           & \phn 6 / 5\\
ANDV9330        & 617.3 & 0.7 & 80 & - / - / $2 \times 2$ & 20 / 60 / 80    & 1200 / 3400 / 3200           &  9 / 6 / 16\\
\hline\vspace*{-2mm}
\end{tabular}
}
\parbox{0.87\textwidth}{Note. --- Continuum intensities $I$ at wavelengths
    $\lambda_0$ for filters with peak transmissions $T$ and exposure times
    $\Delta t$.}
\end{center}
\end{table*}

Three complete data sets with $2 \times 2$ binning ($\sim$0.072\arcsec\
pixel$^{-1}$) were taken on 31~May and 1~June 2012, which included images of the
target \textsf{TG} and pinhole \textsf{PH} mounted in the AO filter wheel at
\textsf{F3} (telescope focal plane). The observations covered neutral iron lines
at 543.4~nm, 557.6~nm, and 617.3~nm. The Fe\,\textsc{i} $\lambda 543.4$~nm line
had already been scanned on 27~May at full spatial resolution (0.036\arcsec\
pixel$^{-1}$) including images of a pinhole array in \textsf{F4} (science
focus at the entrance of the GFPI). Simultaneous \textsf{BBC} images were
recorded (Tab.~\ref{TAB01}). Line scans with the GFPI were usually carried out
using a step width of eight digital-analog (DA) steps, whereas the pre-filter
scans were performed with one DA step. One DA step corresponds to 0.26--0.41~pm
in the spectral range from 530--860~nm.

Broad-band data were taken in the \textsf{BIC} on 26 and 27~May 2012
just before the \textsf{NBC} and \textsf{BBC} measurements. The observing scheme 
was the same for all available pre-filters (396.8~nm, 430.7~nm, 450.6~nm). In addition 
to a few observations of the quiet Sun and a sunspot, images of the target and
pinhole in \textsf{F3}, the pinhole-array in \textsf{F4}, and the pinhole
mounted in the GPU at \textsf{F2} are also included. All data in the \textsf{BIC} 
were taken with a combination of two spare lenses with $f = 500$~mm and
1250~mm because a second lens with $f=500$~mm was not available yet. Thus, the
image scale of 0.0124\arcsec\ pixel$^{-1}$ oversamples the diffraction limit at
these wavelengths by a factor of about three. The proper image scale of
0.032\arcsec\ pixel$^{-1}$ has been obtained in a following campaign 
by an one-to-one imaging.

\subsection{Intensity estimates for the narrow-band channel}

The wavelength $\lambda_0$, FWHM, and transmission $T$ of different \textsf{NBC}
filters are summarized in Tab.~\ref{TAB02} together with the selected binning,
the counts at continuum wavelengths, and frame rates at a given exposure time
$\Delta T$. The results reveal that at full spatial resolution very long
exposure times of up to 100~ms are necessary to achieve at least 2000 counts in
the continuum of most of the measured spectral lines for most of the filters
with $T \sim 40$\% and a $\mathrm{FWHM} \sim 0.3$--0.4 nm. As a consequence, one
can reach only very low frame rates. A $2 \times 2$-pixel binning speeds up the
observations and reduces the exposure times. The situation changes when choosing
filters with higher transmission. The 617.3~nm filter with $T \sim 80$\% and a
$\mathrm{FWHM} \sim 0.74$~nm yields reasonable frame rates and exposure times
even without binning.

\begin{figure}[t]

\includegraphics[width=0.9\columnwidth]{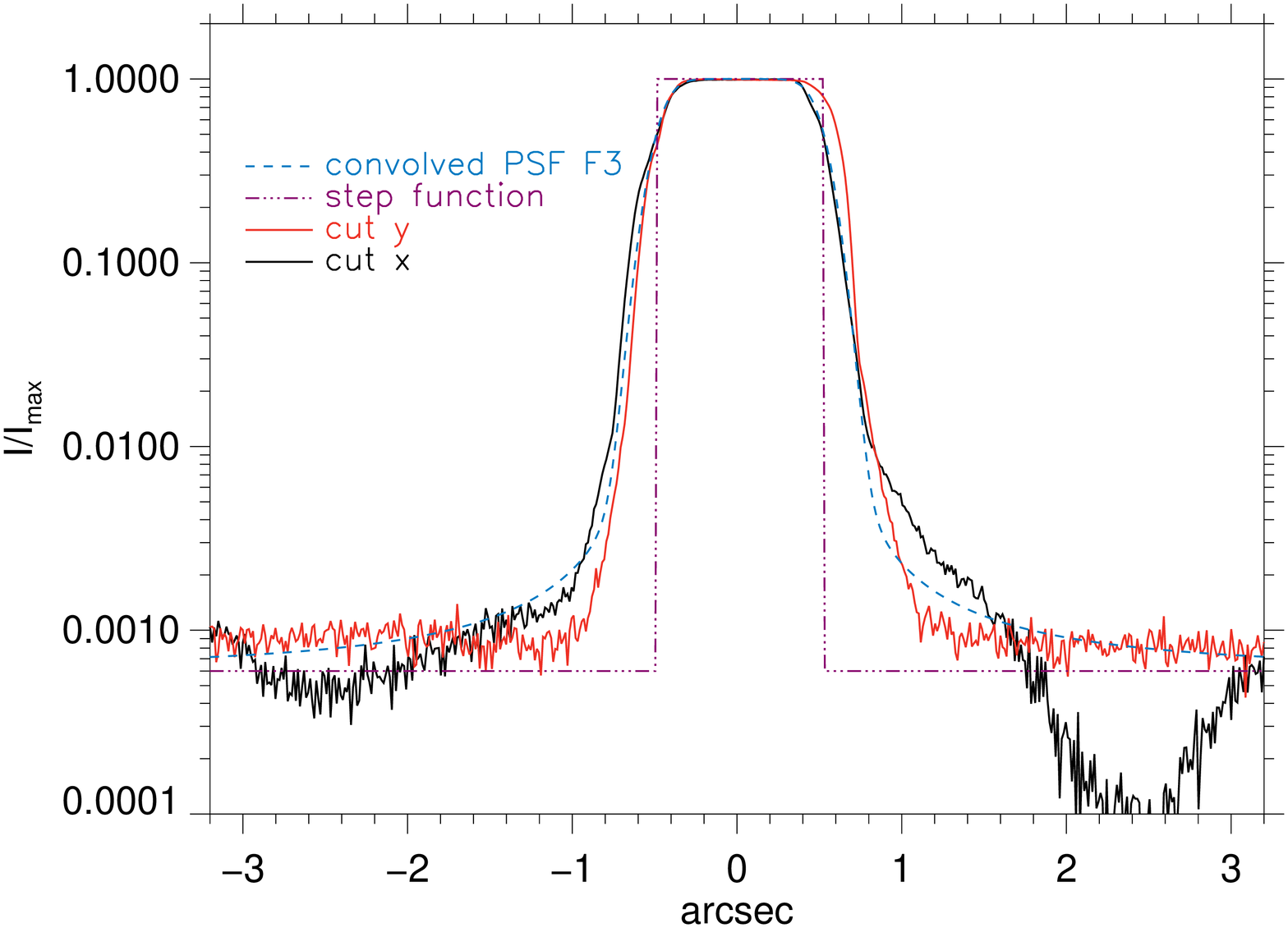}
\hfill
\centering
\includegraphics[width=\columnwidth]{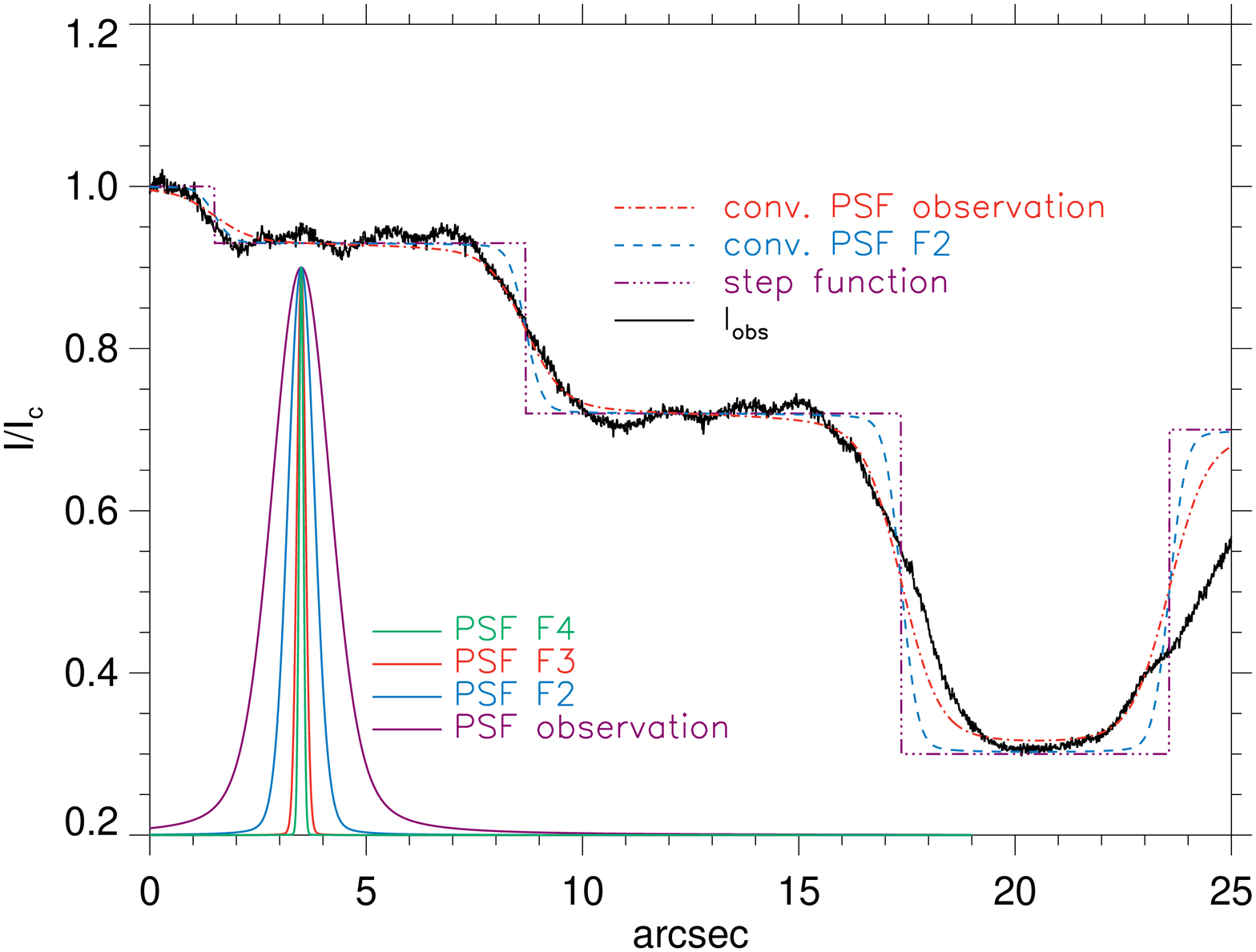}
\caption{Derivation of the PSF estimate at 430~nm for the pinhole in \textsf{F3}
    (\textit{top}) and for the complete optical train from a sunspot
    observation (\textit{bottom}), where the PSF estimates of the four different
    focal planes are displayed in the lower left corner.}
\label{FIG02}
\end{figure}

\subsection{Estimates of the spatial point spread function}

Knowledge of the instrumental point spread function (PSF) provides an estimate
of both the spatial resolution and the \textit{spatial} stray-light level.\cite{2011A&A...535A.129B,
2012A&A...537A..80L} Using a reference such as a pinhole or a blocking edge in
the focal plane, the PSF of all optics downstream can be derived. The combined
PSF of time-variable seeing, telescope, and post-focus instruments, can also be
derived from the observations of, e.g., sunspots with their steep spatial
intensity gradients.

The PSF of the optical train at the GREGOR telescope relevant for the different 
GFPI channels was calculated based on reduced images of
pinholes located in the focal planes \textsf{F4}, \textsf{F3}, \textsf{F2}, and
on sunspot images (see Tab.~\ref{TAB01}). All available calibration images,
e.g., target or pinhole images, were averaged for each wavelength step or image
burst. However, only a single image was selected for solar observations because
of the variable seeing. The pinhole images were normalized
to the maximum intensity $I_{\rm max}$ inside the FOV, 
whereas the sunspot data were normalized to unity in a quiet Sun region. 
All data were taken without real-time correction of the AO
system and consequently correspond to the static performance of the optical
system.

\subsection{Derivation of the point spread function estimates}

To obtain an estimate of the PSF, we took cuts along the $x$- and $y$-axes 
of the CCD across the center of the pinhole in each pinhole observation.  
We defined a step function that is assumed to represent the physical extent of the true
pinhole. The borders of the step function were set to intersect the observed
intensity along the cuts at about the 50\% level (top panel of
Fig.~\ref{FIG02}). We constructed a convolution kernel from a combination of a
Gaussian (variance $\sigma$) and a Lorentzian function (parameter $a$) and
convolved the step function with the kernel. A modification of the parameters
($\sigma, a$) within a specific range yielded finally the kernel that best
matched the convolved step function and the observed intensity along the
horizontal and vertical cuts. The same method was applied to all
pinhole observations (\textsf{F4}, \textsf{F3}, and \textsf{F2}) to derive a PSF
estimate for each focal plane and wavelength. The sunspot observations were
modeled by a similar step function at the two transitions between quiet Sun, penumbra,
and umbra (bottom panel of Fig.~\ref{FIG02}).

The resulting PSF estimates of the four focal planes for the \textsf{BIC} data at
430.7~nm are displayed in the lower left corner of the bottom panel of
Fig.~\ref{FIG02}. The width of the PSF estimates is similar for \textsf{F4} and
\textsf{F3}, where only static optical components contribute to the PSF, but
increases significantly when passing to the focal plane \textsf{F2} that
experiences telescope seeing and seeing fluctuations along the Coud\'e train.
The PSF derived from the sunspot observations includes all seeing effects and
roughly doubles its width relative to the PSF at \textsf{F2}.

\begin{figure}[t]
\centering
\begin{minipage}[c]{0.44\textwidth}
\includegraphics[width=\textwidth]{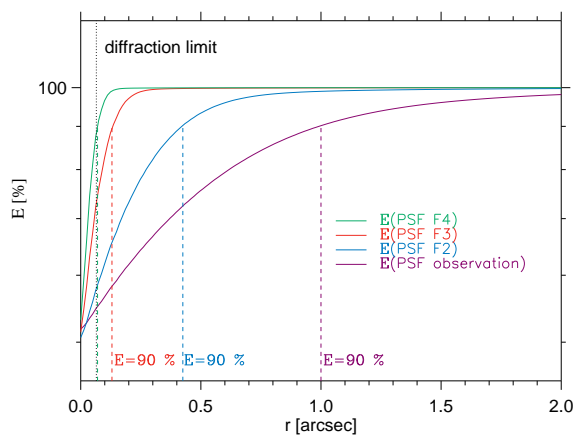}
\end{minipage}
\caption{Energy enclosed within the radius $r$ for the 
    PSF estimates at 430.7~nm of the four different 
    focal planes. The \textit{dashed vertical lines} 
    denote the radius where 90\% of the energy is 
    enclosed. The \textit{dotted vertical line} 
    denotes the diffraction limit at 430.7~nm.}
\label{FIG03}
\end{figure}

PSF estimates for all four focal planes were obtained in case of the
\textsf{BIC} data at 430.7~nm, but only for some of the focal planes in case of the
\textsf{BBC} and \textsf{NBC} data (see Tab.~\ref{TAB03a}). The optical performance and
the expected stray-light level can be best quantified from the total energy
enclosed in a given radius, i.e., a radial integration of the PSF. Figure~\ref{FIG03} shows
the enclosed energy $E$ for the PSF estimates at 430.7~nm. The curves provide two characteristic values,
namely, the energy enclosed at the diffraction limit $E(r_\mathrm{DL})$ and the radius $r(E = 90\%)$ at
which 90\% of the energy is enclosed. The former value gives an estimate of the generic
stray-light when subtracted from 100\%. The latter value can be
used as generic estimate of the spatial resolution. Table~\ref{TAB03a} lists 
the two values for all analyzed wavelengths and channels, i.e., \textsf{BIC}, \textsf{BBC} and \textsf{NBC} at specific wavelengths.

\begin{table}[!t]
\begin{minipage}[c]{0.56\textwidth}
\begin{scriptsize}
\addtolength{\tabcolsep}{-2pt}
\begin{tabular}{lccccccc}
\hline\hline
Channel                       & \multicolumn{4}{c}{BIC $\lambda 430.7$~nm}        & \multicolumn{3}{c}{BBC $\lambda 543.4$~nm}\rule[-1.5mm]{0mm}{4.5mm} \\
\hline 
Sampling                     & \multicolumn{4}{c}{0.0124\arcsec\ pixel$^{-1}$}       & \multicolumn{3}{c}{0.034\arcsec\ pixel$^{-1}$}\rule{0mm}{3mm} \\
                              & \multicolumn{4}{c}{(0.0126\arcsec\ pixel$^{-1}$)}     & \multicolumn{3}{c}{(0.036\arcsec\ pixel$^{-1}$)} \\
Diff.\ limit                  & \multicolumn{4}{c}{0.072\arcsec}                      & \multicolumn{3}{c}{0.091\arcsec}\rule[-1.5mm]{0mm}{3mm} \\
\hline
Focal plane                   & \textsf{F4} & \textsf{F3} & \textsf{F2} & Solar obs.  & \textsf{F4} & \textsf{F3} & Solar obs. \rule[-1.5mm]{0mm}{4.5mm} \\
\hline
$r(E = 90\%)$ $[$\arcsec$]$   & 0.07        &   0.13      &    0.43     & 1.00        & 0.11        &   0.17      & 2.4       \rule{0mm}{3mm} \\
$E(r_\mathrm{DL})$            & 90\%        &   73\%      &    58\%     & 55\%        &  85\%       &   74\%      & 52\%      \rule[-1.5mm]{0mm}{3mm} \\
\hline
\end{tabular}\\
\medskip

\addtolength{\tabcolsep}{-1.5pt}
\begin{tabular}{lccccc}
\hline\hline
Channel                      & \multicolumn{3}{c}{NBC}                    & NBC                   & NBC \rule[-1.5mm]{0mm}{4.5mm} \\
                             & \multicolumn{3}{c}{$\lambda 543.4$~nm}     & $\lambda 557.6$~nm    & $\lambda 617.3$~nm \\
\hline
Sampling                     & \multicolumn{3}{c}{0.034\arcsec\ pixel$^{-1}$}   & 0.069\arcsec\ pixel$^{-1}$ & 0.067\arcsec\ pixel$^{-1}$\rule{0mm}{3mm} \\
                             & \multicolumn{3}{c}{(0.036\arcsec\ pixel$^{-1}$)} & (0.072\arcsec\ pixel$^{-1}$) & (0.072\arcsec\ pixel$^{-1}$) \\
Diff.\ limit                 & \multicolumn{3}{c}{0.091\arcsec}                 & 0.094\arcsec  & 0.103\arcsec \rule[-1.5mm]{0mm}{3mm} \\
\hline
Focal plane                  & \textsf{F4} & \textsf{F3} & Solar obs.           & \textsf{F3}   & \textsf{F3} \rule[-1.5mm]{0mm}{4.5mm} \\
\hline
$r(E = 90\%)$  $[$\arcsec$]$ & 0.12        & 0.20        & 2.65                 & 0.19          & 0.19 \rule{0mm}{3mm} \\
$E(r_\mathrm{DL})$           & 85\%        & 71\%        & 51\%                 & 75\%          & 76\% \rule[-1.5mm]{0mm}{3mm} \\
\hline
\end{tabular}
\end{scriptsize}
\end{minipage}\medskip
\caption{Summary of observations at different wavelengths. 
    Theoretical values are given in parentheses.}
\label{TAB03a}
\end{table}

A comparison between the value of the diffraction limit and the radius where
90\% of the energy is enclosed shows that the optics behind \textsf{F4} performs
close to the diffraction limit. The optics downstream of \textsf{F3} performs
slightly worse with a total enclosed energy of about 70--75\% of the
diffraction-limited case. This implies a spatial stray-light level of about 25\%
created by the optics downstream of \textsf{F3}. The corresponding value at \textsf{F4}
is about 10--15\%. The values of both the stray-light level and $r(E = 90\%)$ experience a 
profound jump when passing to \textsf{F2} and beyond. However we note, that all these data were taken 
at mediocre seeing conditions and without AO correction.

\subsection{Spectral resolution, spectral stray-light, and blue-shift}

\begin{figure}[t]
\centering
\includegraphics[width=\columnwidth]{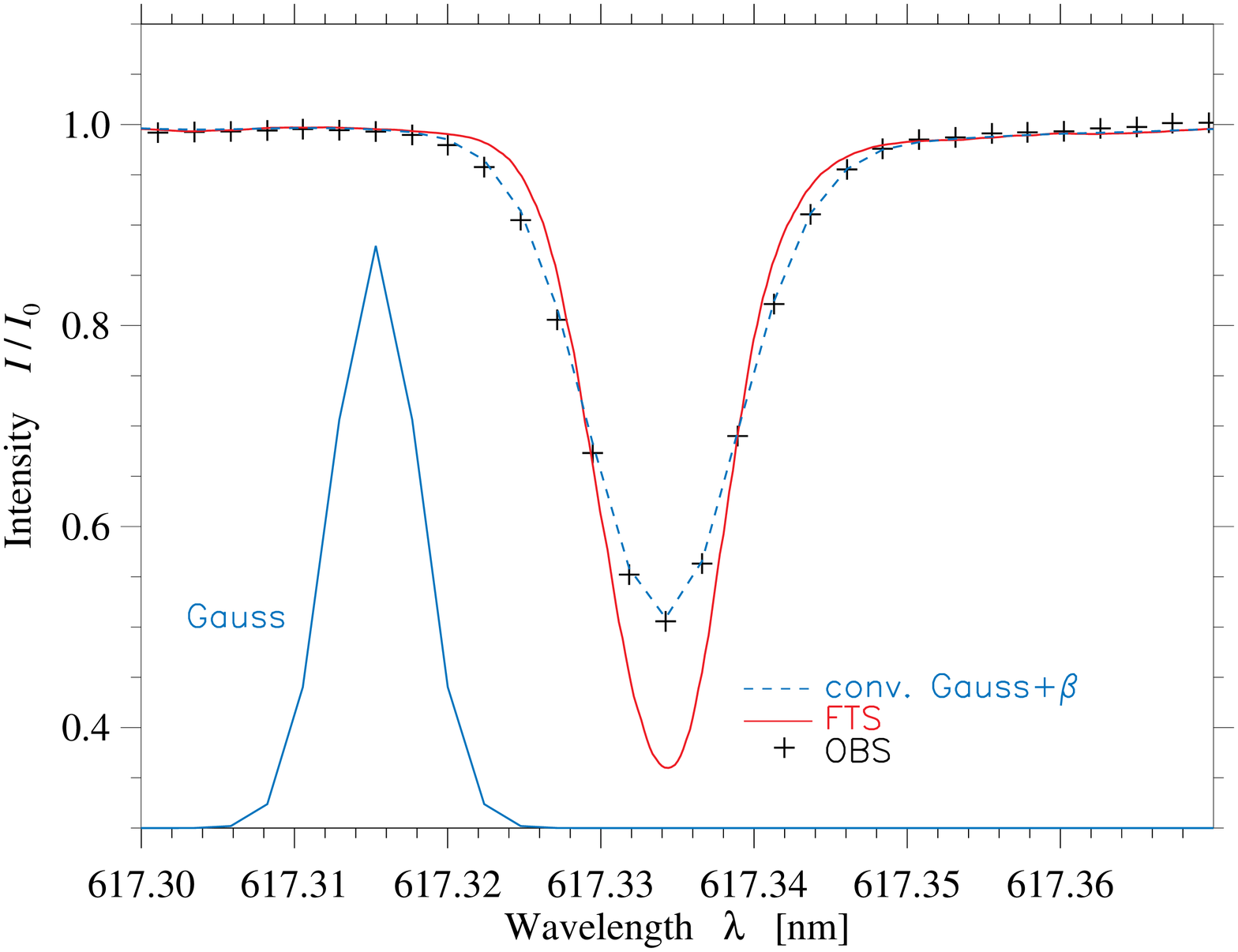}
\hfill
\hspace{-2mm}\includegraphics[angle=90,width=0.96\columnwidth]{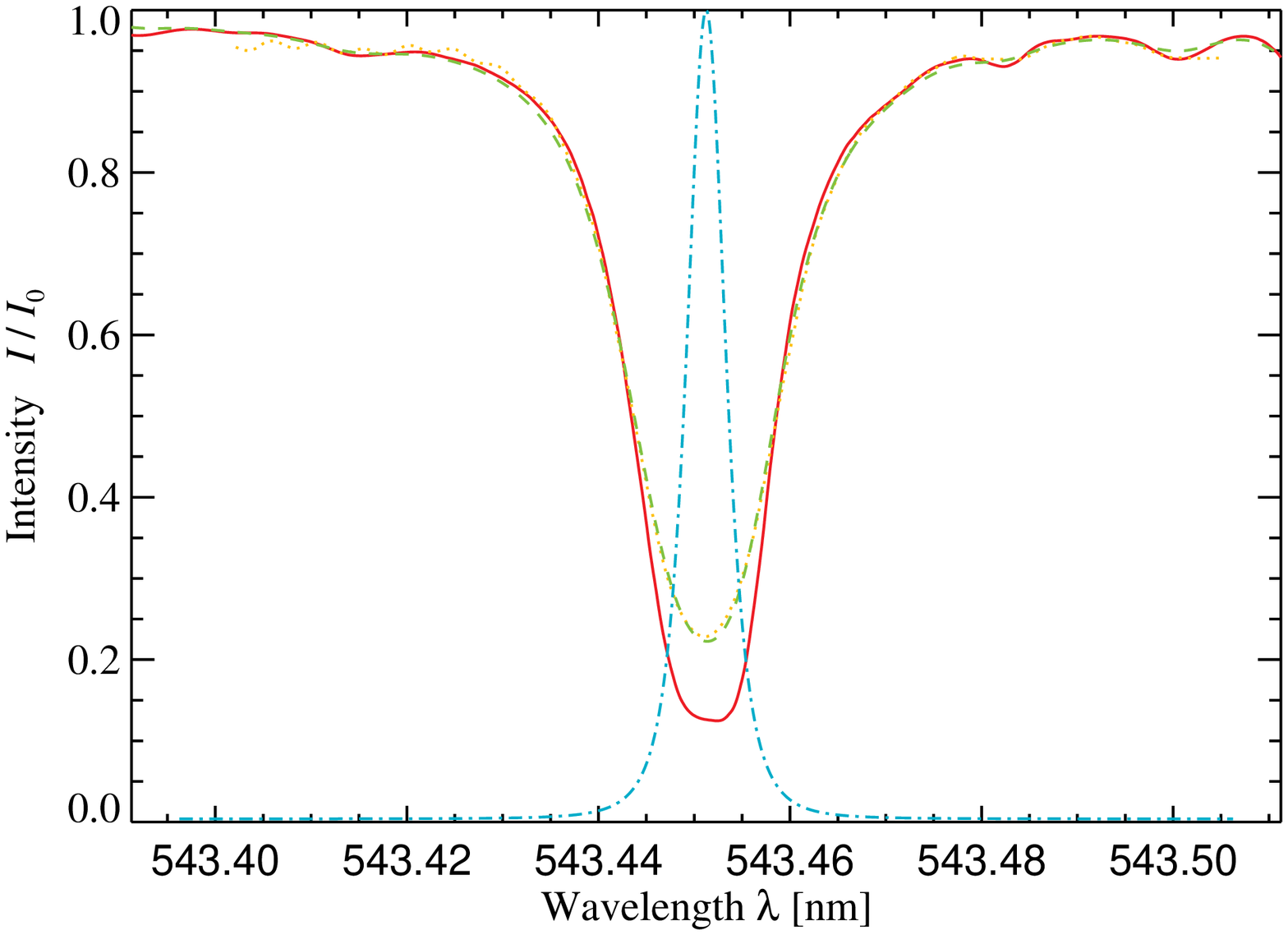}
\caption{\textit{Top:} close-up of the Fe\,\textsc{i} $\lambda 617.3$~nm line -- observed spectrum
    (\textit{black pluses}), FTS atlas spectrum (\textit{red solid line}), and FTS convolved with the best-fit Gaussian kernel (\textit{blue dashed line}). An appropriate stray-light offset ($\beta$) was
    added after the convolution. The Gaussian kernel in the left corner is displayed in arbitrary units. 
    \textit{Bottom:} comparison of averaged observed quiet Sun spectrum (\textit{dotted orange line}) of the Fe\,\textsc{i} $\lambda 543.4$~nm line, FTS atlas spectrum (\textit{solid red line}), and FTS-atlas spectrum convolved with an Airy-function (\textit{dash-dot turquoise line}) obtained from a least-squares fit with the effective finesse and an additional parasitic light contribution as free parameters (\textit{dashed green line}).}
\label{FIG04}
\end{figure}

The spectral resolution and the \textit{spectral} stray-light inside the
\textsf{NBC} was estimated by a convolution of Fourier Transform Spectrograph
(FTS) atlas spectra\cite{kurucz+etal1984, 1999SoPh..184..421N} with a Gaussian of 
width $\sigma$ and a subsequent addition of a constant wavelength-independent stray-light offset
$\beta$.\cite{2004A&A...423.1109A, 2007A&A...475.1067C} This component of
stray-light corresponds to light scattered onto the CCD detector without being
spectrally resolved. Therefore, it changes the line depth of observed spectral
lines. The convolved FTS spectra in each wavelength range were compared with sets of 
spatially averaged observational profiles that either covered the full
pre-filter transmission curve or only the line inside the same range that is
usually recorded in science observations (543.4~nm, 557.6~nm, 617.3~nm). The
upper panel of Fig.~\ref{FIG04} shows the average observed spectrum at
Fe\,\textsc{i} $\lambda 617.3$~nm, the original FTS spectrum, and the FTS
spectrum after the convolution with the best-fit Gaussian kernel and the
addition of the stray-light offset. The method has some ambiguity between
$\sigma$ and $\beta$, which can be modified in opposite directions over some
range near the best-fit values without significantly degrading the reproduction
of the observed spectra. Therefore, the values listed in Tab.~\ref{TAB03} have
an error of about $\pm 0.5$~pm in $\sigma$ and $\pm 5$\% in $\beta$. The
stray-light level $\beta$ inside the \textsf{NBC} is below 10\% and the
spectral resolution is ${\cal R} \sim 100,000$, which is significantly below the theoretically
expected value of ${\cal R} \sim 250,000$. The dispersion values derived from
the observed spectra are listed in Tab.~\ref{TAB03} and correspond to eight DA
steps. In addition, we also confirmed the maximal blueshift induced in the spectra, which is caused by the etalon 
mounting inside a collimated beam. The blueshift was determined from a set of observational flat-field data. 
All resulting numbers for both dispersion and blueshift are close to the
theoretically expected values (see Tab.~\ref{TAB03}).


\begin{table}[t]
\begin{center}
\caption{Spectral characteristics of the GFPI data.}
\label{TAB03}
\footnotesize
\medskip
\addtolength{\tabcolsep}{-2pt}
\begin{tabular}{lccccc}
\hline\hline
Wavelength                     && 543.4                  &    557.6                  &   617.3                   &   \phn\phn 630.25\cite{2011A&A...533A..21P}\rule{0mm}{4mm} \\
\rule[-2mm]{0mm}{4mm}          && [nm]                   &     [nm]                  &   [nm]                    & \\
\hline
Dispersion (8~DA) [pm]         &&   2.08                 &    2.15                   &   2.36                    &   2.31 \rule{0mm}{4mm}                      \\
                               &&  (2.09)                &    (2.15)                 &   (2.36)                  &   (2.41)                                    \\
Dispersion (1~DA) [pm]         &&   --                   &      --                   &     --                    &    --                                        \\                           
                               &&  (0.261)               &   (0.268)                 &   (0.296)                 &                                              \\
$\sigma$          [pm]         &&   2.62                 &    2.07                   &   2.81                    &   1.65                                       \\
$\beta$           [\%]         &&   6.6                  &    6.5                    &   7.8                     &   14                                         \\
$\lambda/\sigma$   --          &&   207200               &    269500                 &   219700                  &   381800 \rule[-2mm]{0mm}{4mm}               \\
$\lambda/\Delta\lambda$  --    &&   \phn 88200           &    114700                 &   \phn 93500              &   162500                                     \\
Max.\ blueshift  [pm]         &&  3.91                  &    3.96                   &  4.49                     &   --\\          
                               &&  (3.72)                &    (3.82)                 & (4.23)                    &   (4.32)\rule[-2mm]{0mm}{4mm}                \\          
\hline\vspace*{-2mm}
\end{tabular}
\parbox{0.81\textwidth}{Note. --- Theoretical values are indicated in parentheses.}
\end{center}
\end{table}

Because of the surprisingly low spectral resolution, we spent substantial effort to clarify the source of this problem. We tried to match FTS spectra convolved with the Airy-function, resulting from a theoretical simulation based on a least-squares fit with the effective finesse and an additional parasitic light contribution as free parameters, with observed profiles from campaigns performed in July and August 2012. It turned out that a finesse of $\sim$\,16  instead of $\sim$\,46 and an additional parasitic light contribution of $\sim$\,7\% are needed to fit the observed profile of, e.g., the 543.4\,nm line (lower panel of Fig.~\ref{FIG04}). 

     A finesse of 16 is in clear contradiction to the results obtained from the finesse-adjustment by laser 
     light that is performed every day at the beginning of each observational run. These adjustments revealed always 
     a finesse of 40-50 for both etalons. Furthermore, a better spectral resolution using the same etalons was 
     achieved at the VTT (${\cal R} \sim 160,000$).\cite{2011A&A...533A..21P} Several possibilities which might 
     explain these observations were explored, e.g., the variation of the finesse between the center and 
     periphery of the etalon plates.\cite{2005PASP..117.1435D} However, none of the additional tests were conclusive. 
     
     A precise re-alignment of the GFPI in 2013 will show, if a stricter requirement for the collimation of the beam through the etalons together with an adjustment of the parallelism of the etalons with a wider laser beam will result in a higher finesse. Additionally we will carry out tests in an optical laboratory to quantify cavity errors caused by, e.g., the micro-roughness of the etalon surfaces.\cite{2006A&A...447.1111S, 2010A&A...515A..85R, 2011A&A...534A..45S} 

\begin{figure*}[!ht]
\centering
\includegraphics[angle=90,width=\textwidth]{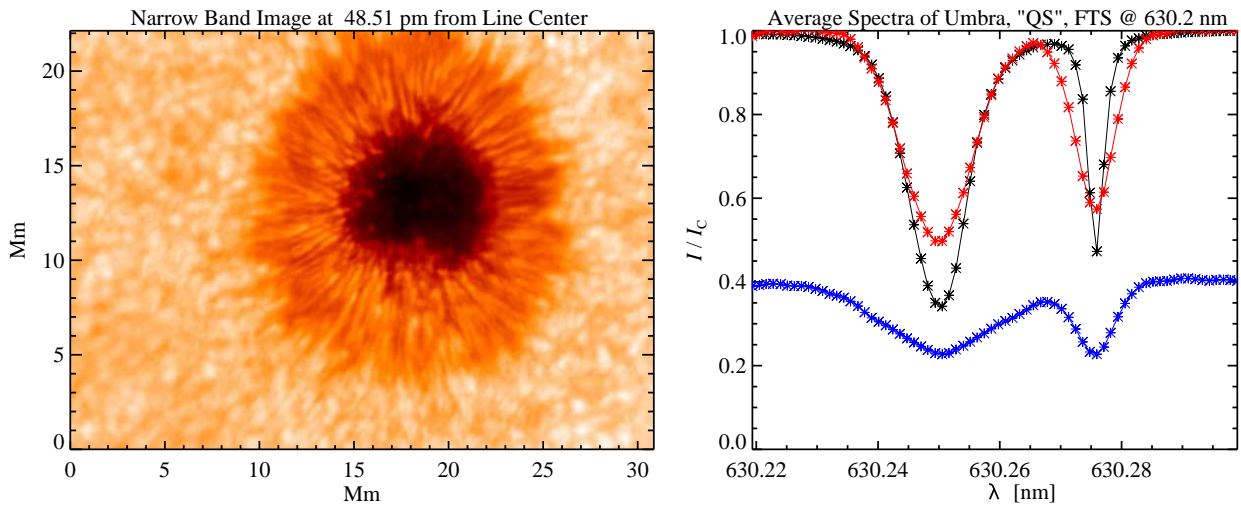}
\caption{Left panel: Deconvolved continuum narrow-band image of the active region NOAA 11538
         at 48.5~pm distance from the line center of the Fe\,\textsc{i} $\lambda 630.25$~nm line. Right panel: 
         measured average profiles of quiet sun (red), umbra (blue), and the FTS-atlas profile
         for comparison. The stars indicate the sampled spectral positions. 
\href{http://www.aip.de/Members/kgpuschmann/OE-AstroPh?set_language=en}{Movie~1}
 shows the entire line-scan, i.e., the deconvolved narrow-band images (left panel) at the consecutive spectral positions (right panel) in steps of 1.20~pm.}
\label{FIG04a}
\end{figure*}

\begin{figure*}[!ht]
\centering
\includegraphics[width=\textwidth]{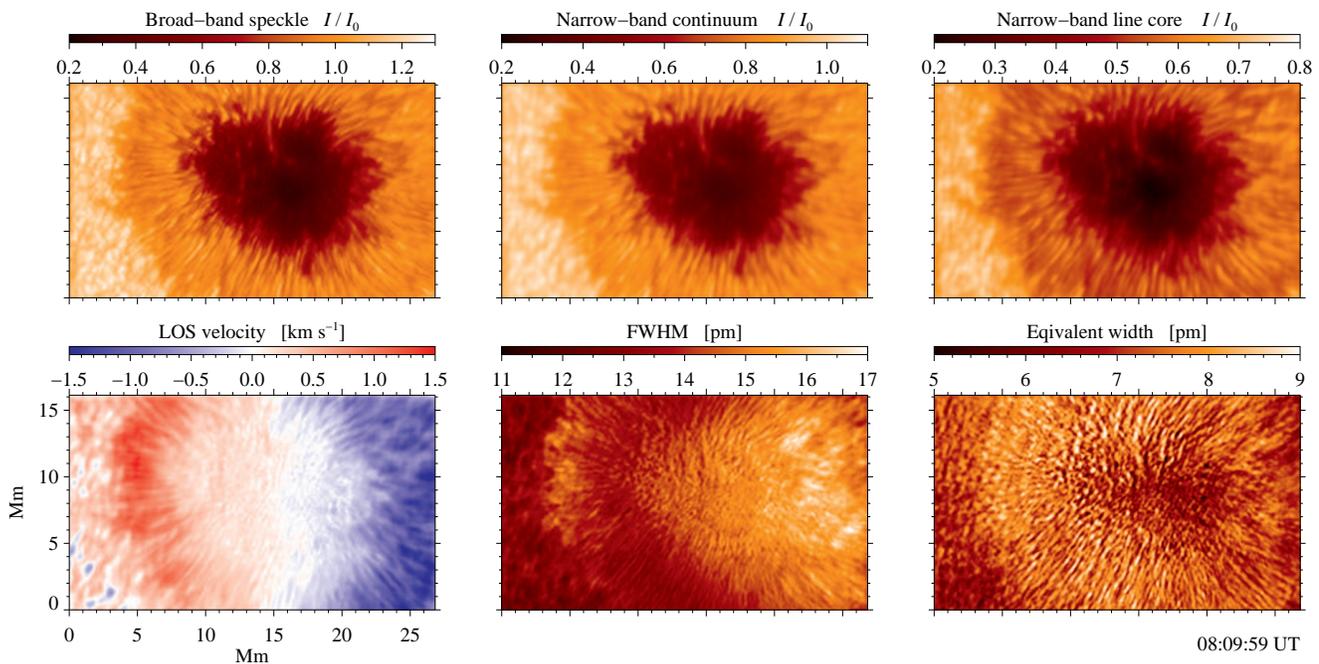}
\caption{Snapshot of a 22-min time-series of the active region NOAA 11530. Upper panels: speckle-reconstructed broad-band image at 620\,nm (left), speckle-deconvolved continuum narrow-band image
         (middle), and line-core narrow-band image (right) of the 
         Fe\,\textsc{i} $\lambda 617.3$~nm line. Bottom panels: 
         line-core velocity (left), full-width at half-maximum (middle), 
         and equivalent width (right). 
         \href{http://www.aip.de/Members/kgpuschmann/OE-AstroPh?set_language=en}{Movie~2} shows the temporal evolution of the individual parameters.}
\label{FIG04b}
\end{figure*}

\subsection{First observations supported by adaptive optics}

First observations with the GFPI at GREGOR were obtained between 28 July and 7 August 2012 with real-time correction of wavefront deformations provided by GAOS. Several data sets covering different spectral
lines (e.g., 543.4~nm, 617.3~nm, 630.2~nm), and different objects and disk positions have been 
reduced and analyzed during the last months.

Figure~\ref{FIG04a} and \href{http://www.aip.de/Members/kgpuschmann/OE-AstroPh?set_language=en}{Movie~1} 
show a detailed example of two-dimensional (2D) spectral line scans with the GFPI 
that were observed on 7 August 2012. The narrow-band continuum image around 630.2~nm is displayed in 
the left panel of Fig.~\ref{FIG04a}. The spectral scan covered the Fe\,\textsc{i} $\lambda 630.25$~nm line and the
nearby teluric O$_2$ line with 70 steps in total and a step size of  $\Delta\lambda= 1.20$~pm
(right panel of Fig.~\ref{FIG04a}). \href{http://www.aip.de/Members/kgpuschmann/OE-AstroPh?set_language=en}{Movie~1}
 shows all narrow-band images (left panel) at the consecutive spectral positions denoted by stars (right panel). The penumbral finestructure and individual umbral dots of the active region NOAA 11538 located at S22 W24 can be clearly identified in the spectral data. While throughout the solar line a significant change of structure and intensity contrast with height in the solar atmosphere is visible, the narrow-band images of the O$_2$ line just reflect the continuum intensity because of the teluric origin of the latter.

On 31 July, in total a 22-min time-series of the active region NOAA 11530 located at S24 W18  (heliocentric angle $\theta=30^{\circ}$) was obtained 
between 8:04 and 8:26 UT. Because the angle between solar south and the center of symmetry was 33.8$^{\circ}$, the direction 
from center to limb corresponds to the $x$-axis of the displayed images. The spectral scan covered the Fe\,\textsc{i} $\lambda 617.3$~nm line with 
86 steps of $\Delta\lambda= 1.17$~pm step size.

\begin{figure*}[!ht]
\centerline{\includegraphics[width=\textwidth]{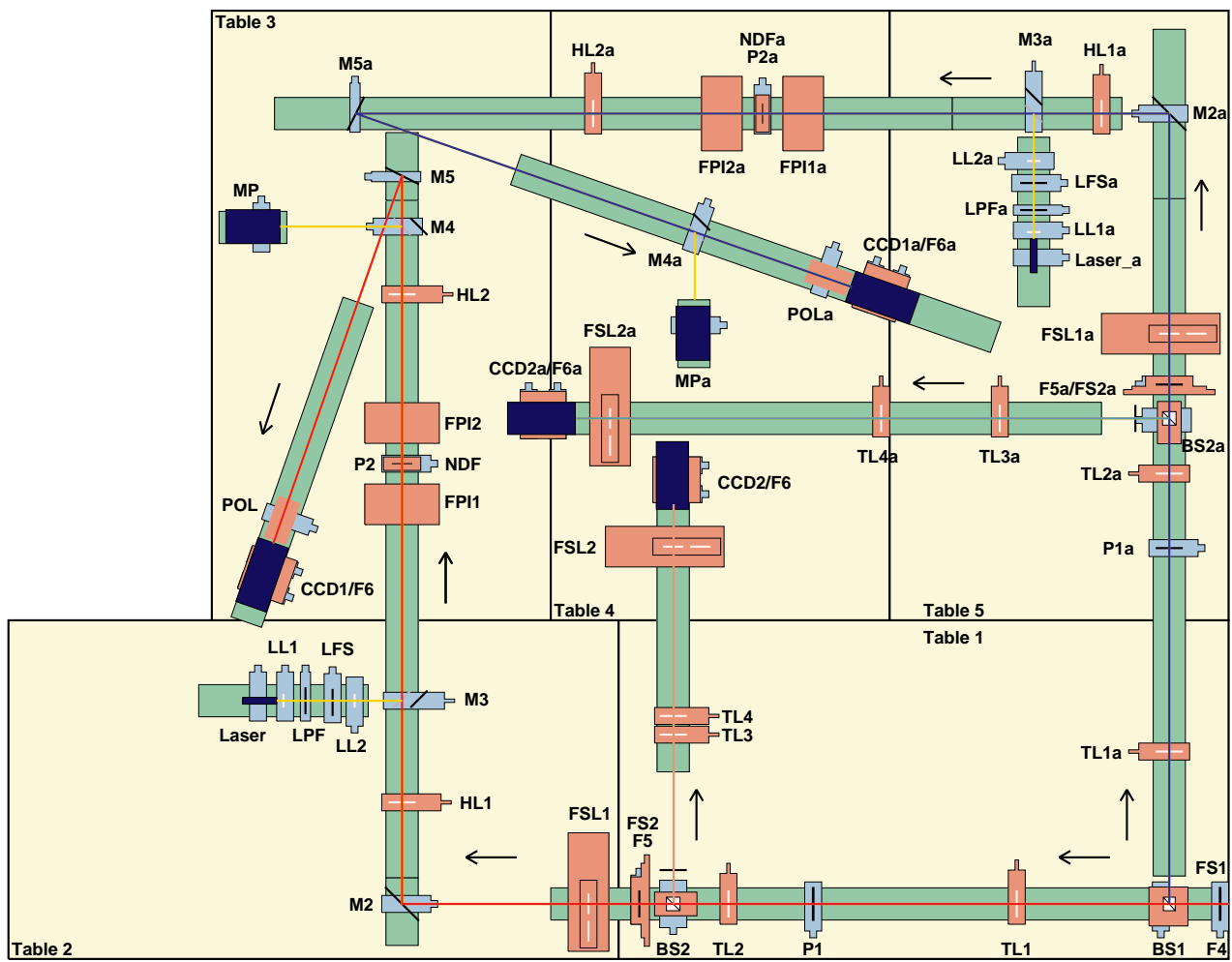}}
\caption{Up-to-scale drawing for the integration of BLISS into the GFPI system.
    The design is based on the study presented in Fig.~\ref{FIG07}. The optical
    elements of BLISS are labeled with an extra `a'. Apart from \textsf{TL2a},
    \textsf{HL2a}, and \textsf{TL4a}, they are identical to those of the GFPI. The \textsf{NBC}
    of GFPI and BLISS is denoted by a red and blue beam, respectively. For the BBC of both instruments, 
    a salmon (GFPI) and turquoise (BLISS) beam distinguish between the two instruments. Laser/Photomultiplier 
    channels of both instruments are highlighted by a yellow beam. The light from the telescope enters at 
    the science focus \textsf{F4}. Black arrows denote the light direction inside both instruments.}  
\label{FIG06}
\end{figure*}

Figure~\ref{FIG04b} shows 2D maps of individual line parameters, i.e., normalized continuum intensity, 
line-core intensity, line-of-sight velocity, full-width at half-maximum, and equivalent width together with the speckle-reconstructed 
broad-band image at 620~nm for one spectral scan. The temporal evolution of these line parameters over the entire 
time-series is presented in \href{http://www.aip.de/Members/kgpuschmann/OE-AstroPh?set_language=en}{Movie~2}. In the LOS velocity map, the thinnest radial filamentary
structures correspond to areas with almost zero velocity (spines), whereas the flow filaments on both center and limb
sides are significantly wider (intraspines). Even with some fluctuations in the final image quality, the evolution of small-scale
structures can be well followed, e.g., the inward motion of penumbral grains. The cadence of the observations is high enough 
to trace nearly continuously the temporal evolution.

The restored broad- and narrow-band images presented here have been obtained by a speckle 
reconstruction\cite{1993PhDT.......155D, 2006A&A...454.1011P} of the broad-band raw data and a subsequent 
speckle deconvolution\cite{1992A&A...261..321K, 2003PhDT.........2J} of the narrow-band raw data. 
In the case of time series, the temporal evolution of line parameters in 2D-maps has been obtained after 
additional data reduction, e.g., image de-rotation/correlation and sub-sonic filtering.\cite{2012AN....333..880P}
In summary, these first observations with the GFPI and GAOS are very promising and suggest that high-cadence
2D spectroscopic observations near the diffraction limit are very likely to be obtained in coming
observing campaigns at GREGOR.

%
%

\section{BLue Imaging Solar Spectrometer}
\label{BLISS}

The spatial resolution of a telescope scales inversely proportional with the
observed wavelength. Therefore, observations at short wavelengths (below 530~nm)
offer the opportunity to obtain data with higher spatial resolution. There are
relatively few ground-based instruments for spectral observations in the blue
spectral region.\cite{1993ApJ...414..345L, 2002A&A...389. 1020R,
2005A&A...437.1159B, 2008A&A...479..213B} All of these observations were carried
out with slit-spectrographs that permit only limited improvements by post-factum
restoration techniques,\cite{2011A&A...535A.129B} which are however applicable
to Fabry-P\'erot-based imaging spectrometers. This motivated the design of
BLISS, which will supplant the blue imaging channel of the GFPI in the near
future.

\subsection{Spectral lines in the blue part of the visible spectrum}

\begin{figure*}[!ht]
\begin{center}
\includegraphics[width=\textwidth]{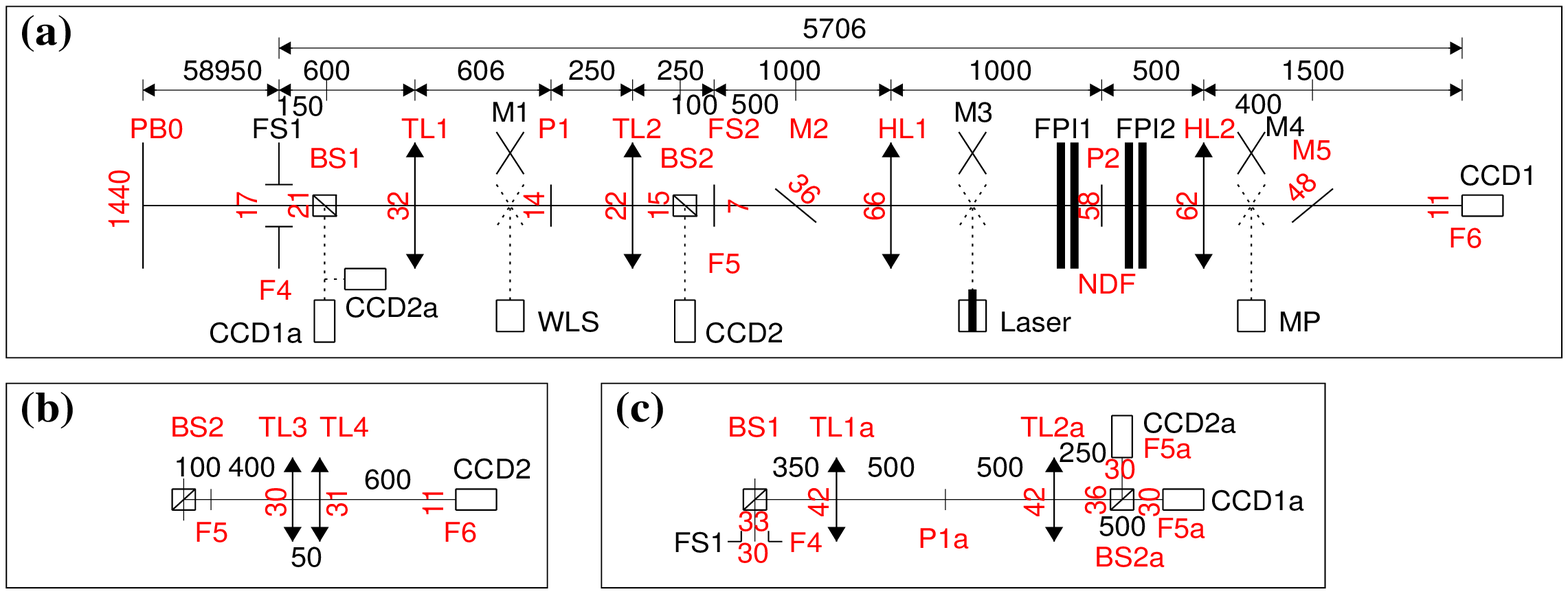}

$ $\\
$ $\\

\includegraphics[width=\textwidth]{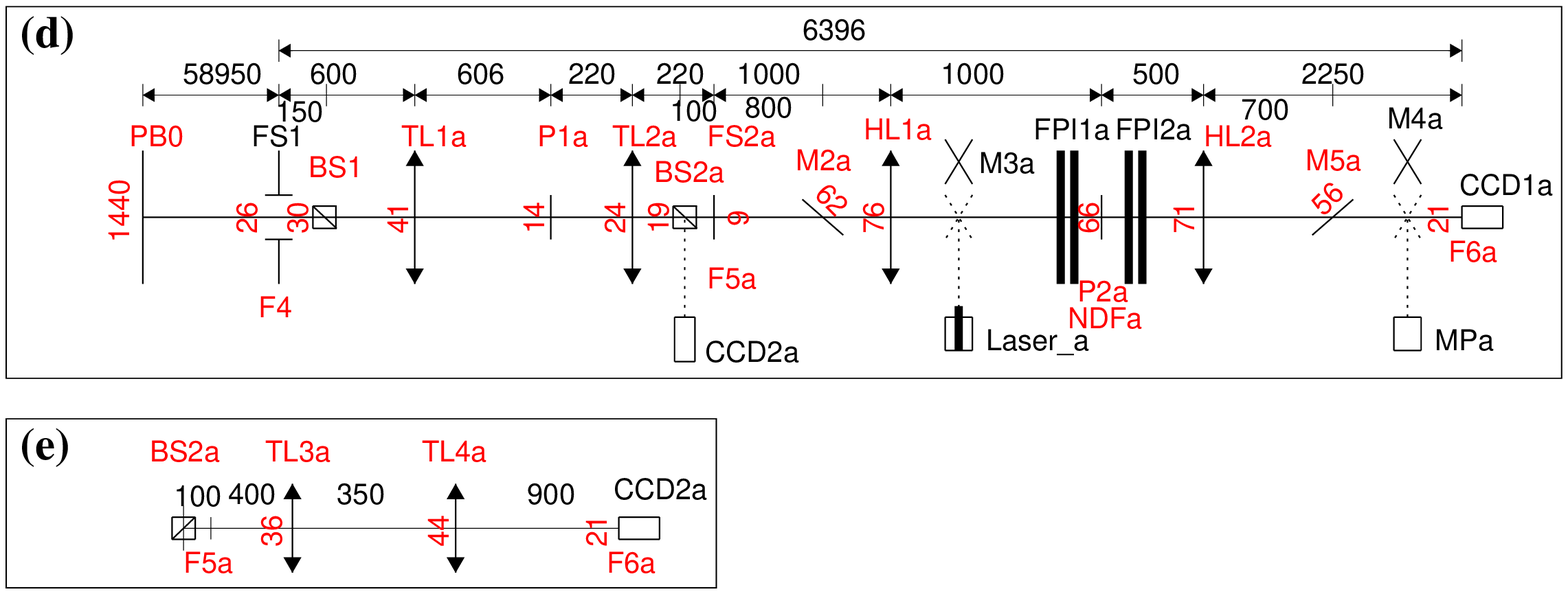}
\end{center}
\caption{Schematical optical design of the current GFPI narrow-band channel 
    \textsf{NBC} \textit{(a)}, broad-band channel \textsf{BBC} \textit{(b)}, and blue imaging 
    channel \textsf{BIC} \textit{(c)}. In the future, \textsf{NBC} \textit{(d)} and \textsf{BBC}
    \textit{(e)} of BLISS will replace the \textsf{BIC} of the GFPI. The diameter of 
    the beam at all optical surfaces, the focal length of the lenses, and the 
    positions of other optical elements are given in millimeters. The optical
    elements of BIC and BLISS are labeled with an extra `a'. The achromatic lenses
    \textsf{TL2a}, \textsf{HL2a}, and \textsf{TL4a} of BLISS have a focal length of 
     220~mm, 2250~mm, and 900~mm, respectively. The mirrors \textsf{M2} and \textsf{M5} of 
     the GFPI will be moved for the integration of BLISS by 200~mm and 50~mm, respectively.}
\label{FIG07}
\end{figure*}

In the wavelength range covered by BLISS, there are several spectral lines and
two molecular bands (G-band $\lambda 430$~nm and CN band-head $\lambda 388$~nm)
of high scientific interest. All Balmer-lines except H$\alpha$ are at shorter
wavelengths than 530~nm, and the Ca\,\textsc{ii}\,H and K lines are the strongest
lines in the visible part of the solar spectrum probing the chromosphere.
Several photospheric resonance lines such as Mg\,\textsc{i} $\lambda 457$~nm,
Sr\,\textsc{ii} $\lambda 407$~nm, and Ba\,\textsc{ii} $\lambda 455$~nm are found
in blue part of the visible spectrum, which also contains several magnetically
insensitive lines\cite{1970SoPh...12...66S} with $g_\mathrm{eff} = 0$ and lines
exhibiting a Zeeman-triplet with splitting factors of $g_\mathrm{eff} = 2.5$ and
higher.\cite{1973SoPh...28....9H} A special case is the pair Fe\,\textsc{i}
$\lambda 461.32$~nm ($g_\mathrm{eff} = 0$) and Cr\,\textsc{i} $\lambda
461.37$~nm ($g_\mathrm{eff} = 2.5$) that can be recorded at the same time.

\subsection{Optical design}

\begin{figure*}[!ht]
\centerline{\includegraphics[width=\textwidth]{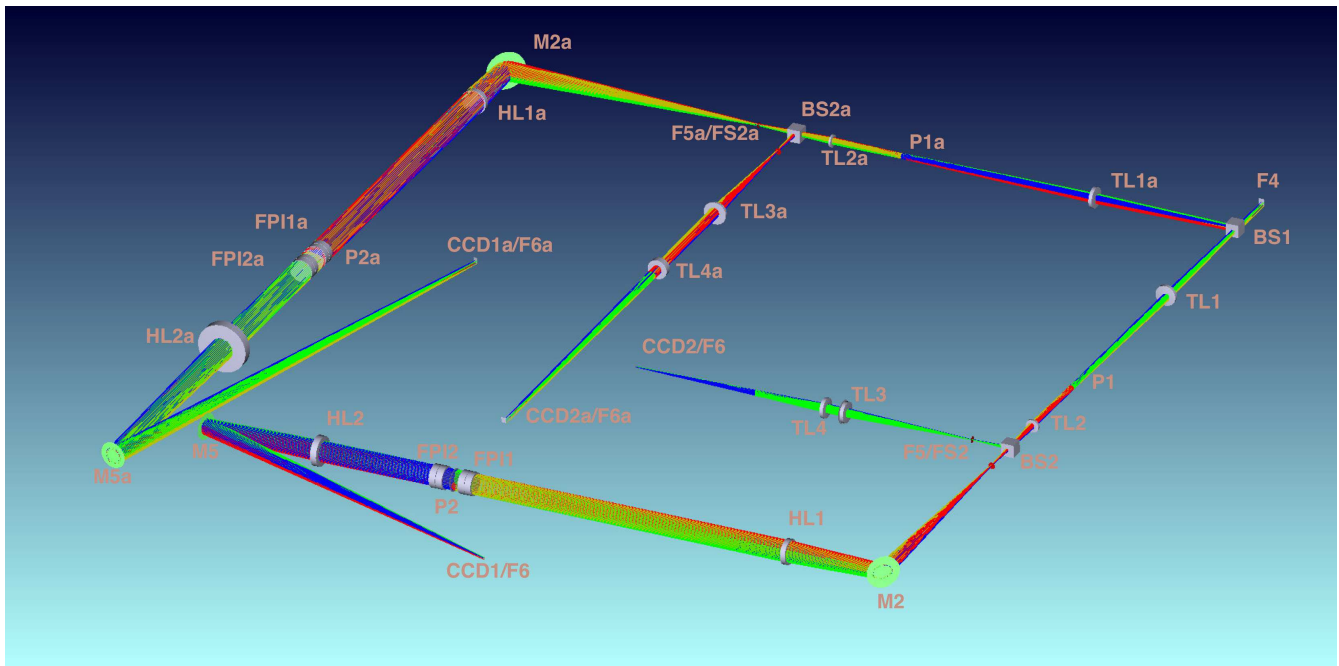}}
\medskip
\caption{Arrangement of GFPI and BLISS in a ZEMAX multi-configuration
    file in shaded modeling for the respective central wave length of each
    instrument and their maximal field dimension. The design confirms the
    calculations in geometrical optics presented in Figs.~\ref{FIG06} and
    \ref{FIG07}. The optical elements of BLISS are labeled with an extra `a'.
    The light coming from the telescope enters at the science focus \textsf{F4}.}
\label{FIG08}
\end{figure*}

The geometrical design of BLISS and its integration into the GFPI is depicted in
Fig.~\ref{FIG06}. Most of the optical components and their labels are identical
to those in Fig.~\ref{FIG01}. The calculations for this initial optical design
revealed that a pupil \textsf{P2} of about 70~mm diameter will be sufficient for
an acceptable maximal blue-shift over the entire wavelength range at a given
image scale and FOV. Thus, the design of BLISS only requires different camera
lenses \textsf{HL2a} and \textsf{TL4a} in the \textsf{NBC} and \textsf{BBC} to
obtain an adequate image scale. The reduction of the diameter of 
\textsf{P2}, caused by several 
modifications of the telescope since 2007, is compensated in the design of 
BLISS. The schematic optical designs of both
instruments are compared in Fig.~\ref{FIG07}. The diameter of the light beam at
the different foci, pupils, and on the relevant optical surfaces has been
calculated by means of geometrical optics and confirmed by a ZEMAX ray-tracing as
for the GFPI.\cite{2007msfa.conf...45P} The ZEMAX ray-tracing is
presented in Fig.~\ref{FIG08}.

A new type of cameras has been envisaged for BLISS, namely, two sCMOS-cameras
from PCO (pco.edge). These cameras have $2560 \times 2160$ pixels with a pixel
size of 6.5~$\mu$m $\times$ 6.5~$\mu$m similar to the GFPI Imager QE cameras.
Using a focal length of $f = 2250$~mm for \textsf{HL2} and $f = 900$~mm for
\textsf{TL4} yields an image scale of 0.028\arcsec\ pixel$^{-1}$ and a FOV of
$71\arcsec \times 60\arcsec$ on both cameras. With this configuration, the
maximal blue-shift across the FOV amounts to 4.4~pm and 6.2~pm at 380~nm and
530~nm, respectively.

In addition, we checked the possibility of interchanging the sCMOS and Imager QE
cameras between the GFPI and BLISS. The almost identical pixel sizes of both
cameras facilitates this task. For the integration of the sCMOS cameras into the
GFPI, a circular FOV with a diameter of $d = 79\arcsec$ and an image scale of
0.036\arcsec\ pixel$^{-1}$ for spectroscopy would result in a maximal blueshift
of 6.9~pm at 630~nm. However, the FOV for polarimetry is limited by the
full-Stokes polarimeter and would remain at its previous size of $25\arcsec
\times 38\arcsec$ unless an upgrade of the polarimeter will be undertaken 
in addition (see Section \ref{polarimetry}). The frame rate at full resolution 
would increase from 7 to 40~Hz, which perfectly suits this observing mode.

The integration of the Imager QE cameras into BLISS would result in an image
scale of 0.027\arcsec\ pixel$^{-1}$ and a FOV of $38\arcsec \times
29\arcsec$ with a maximum blueshift of 1.1~pm and 1.6~pm at 380~nm and
530~nm, respectively. The lower frame rates would match the longer exposure
times at shorter wavelength. A super-achromatic optical setup for BLISS -- as
also foreseen for the GFPI -- would be preferable but otherwise all lenses
except \textsf{TL4a} of BLISS could be purchased as off-the-shelf achromats.

The distribution of the two instruments on the five optical tables in the GREGOR
observing room is shown in Fig.~\ref{FIG06}. All optical elements of BLISS have
been labeled with an additional ``a'' to distinguish them from those of the
GFPI. Behind the common dichroic beamsplitter cube \textsf{BS1}, each instrument
has its own \textsf{NBC} and \textsf{BBC}. A displacement of the folding mirrors
\textsf{M2} and \textsf{M5} of the GFPI to a distance of 700~mm and 350~mm from
\textsf{F5} and \textsf{HL2}, respectively, yields some free space on optical
table 3 that can be used for the \textsf{NBC} of BLISS. The ZEMAX ray-tracing
revealed changes in the optical path when considering the etalons plates in the
design. Thus, the GFPI \textsf{NBC} has to be shortened further by reducing the
distance between \textsf{P2} and \textsf{HL2} from 500~mm to 350~mm. This detail
is not considered in Figs.~\ref{FIG06} and \ref{FIG07}.

The beam in the \textsf{NBC} of BLISS is also folded twice by \textsf{M2a} and
\textsf{M5a} at a distance of 800~mm and 700~mm from \textsf{F5a} and
\textsf{HL2a}, respectively. In the \textsf{BBC} of BLISS, \textsf{TL3a} and
\textsf{TL4a} are separated by 350~mm, in contrast to the 50~mm between
\textsf{TL3} and \textsf{TL4} in case of the GFPI. Two computer-controlled
filter sliders \textsf{FSL1a} and \textsf{FSL2a} will again switch between two
sets of interference filters. BLISS is mainly designed for spectroscopy because
of the expected low photon numbers in the blue spectral region. Nevertheless, a
full-Stokes polarimeter can easily be integrated into the system. As in case of
the GFPI, a laser/photo-multiplier channel for finesse adjustment of the etalons
will be implemented. The white-light channel of the GFPI will be removed and be
replaced by an external white-light source shared by all post-focus
instruments.

\subsection{Camera system}\label{SEC05_3}

Modern sCMOS cameras are capable of delivering high frame rates using
large-format sensors with low readout noise, which makes them ideally suited for
an application in solar physics. The pco.edge is a potential candidate for
BLISS. The camera has a full well capacity of 30,000~e$^{-}$ and uses 16-bit
digitization.  The camera has a maximum frame rate of 40~Hz when operated in the
global shutter mode. The cameras are running in a ``fast scan mode'' (286~MHz)
converting the 16-bit signal internally to 12-bit because of speed
limitations of the camera link interface. The signal losses due to the
compression are roughly a factor of ten smaller than the shot noise of the
camera signal. The readout noise is in the order of 2.3~e$^{-}$. The dark
current consists of a part related to the exposure time, i.e., 2--6~e$^{-}$
pixel$^{-1}$ s$^{-1}$ and a part related to the sensor readout time, which is
constant for a given pixel clock, i.e., 0.6 e$^{-}$ pixel$^{-1}$. Peltier
cooling of the sensor ensures an operating temperature of $+5^{\circ}$~C. The
camera has a quantum efficiency of $\sim$30\% and $\sim$54\% at 380~nm and
530~nm, respectively, similar to the Imager QE cameras currently used in the
GFPI ($\sim$36\% and $\sim$60\%).

To handle the extremely large data acquisition rate of about 300~MB~s$^{-1}$ at
frame rates of 40~Hz, each camera will be controlled by a separate PC, in
which the data will be stored on local RAID~0 systems. The integration of the
cameras in the control software is straightforward because the DaVis software
will be operational on the new system with minor changes only. We will
  perform first in-situ test-measurements with these cameras during an
  upcoming GFPI-technical campaign in April 2013 at the GREGOR telescope, 
  whose results will undergo subsequently a careful and detailed analysis.

However, the feasibility of using the pco.edge for BLISS has still to be
demonstrated. A detailed evaluation of the photon statistic in the wavelength
range 380--530~nm will help with the final decision. Rather long exposure times
can be expected for the blue wavelength range and the exposure time of the
cameras currently has an upper limit of 100~ms due to a relatively high dark
current in the global shutter mode. On the other hand, high frame rates would be
extremely beneficial when operating the GFPI in the vector polarimetric mode,
because at present this instrument is limited to just 5--7~Hz at full resolution. 
The almost identical pixel size facilitates interchanging the cameras between 
the two instruments.

\begin{figure}[t]
\includegraphics[width=\columnwidth]{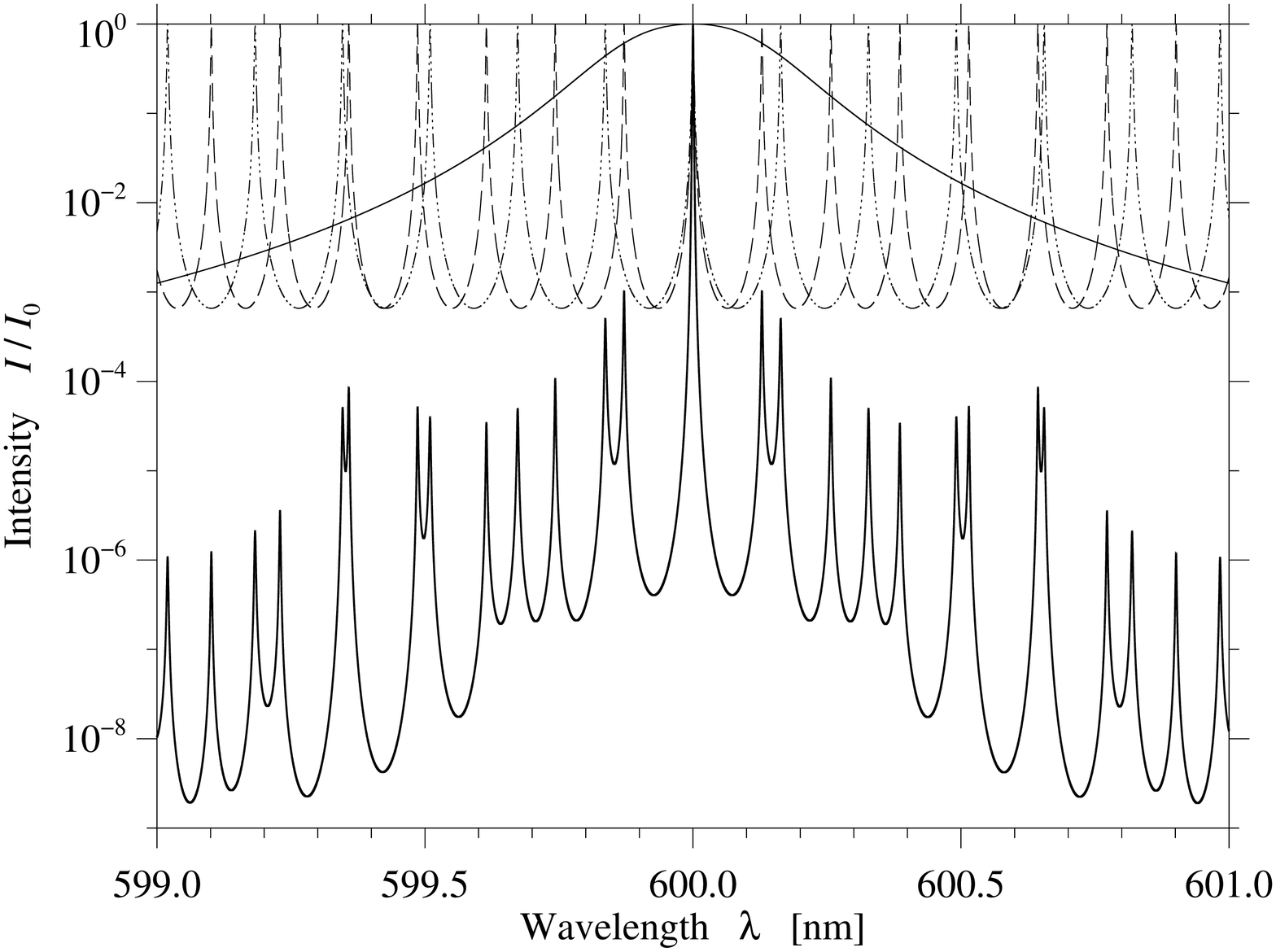}
\hfill
\includegraphics[width=\columnwidth]{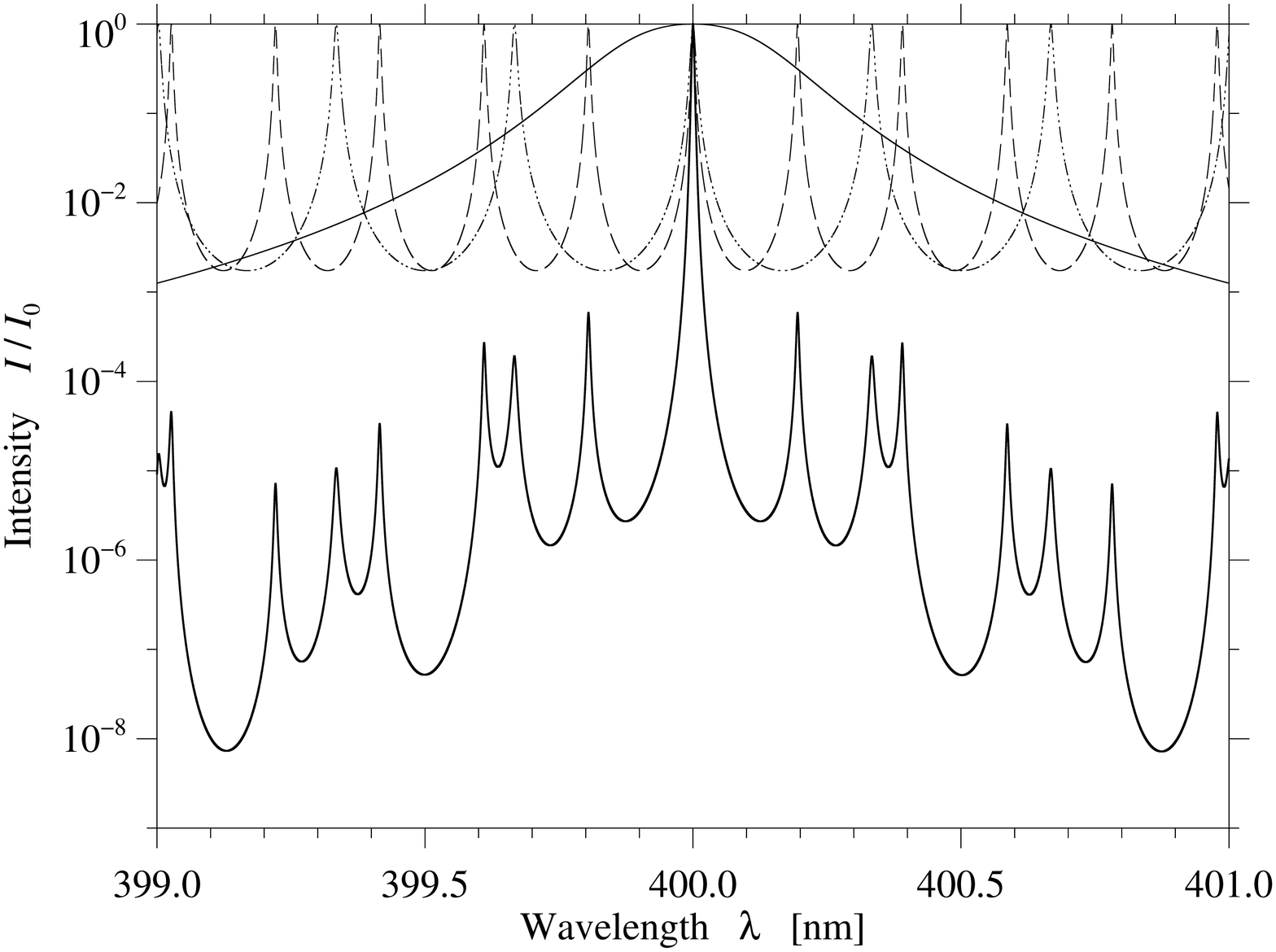}
\caption{Comparison of GFPI (\textit{top}) and BLISS (\textit{bottom})
    transmission profiles. The curves correspond to a narrow-band interference
    filter ($\mathrm{FWHM} = 0.3$~nm, \textit{thin solid line}), etalon~1 ($R =
    0.95/0.92$, $d = 1.4/0.41$~mm, $\mathrm{FWHM} = 2.1/5.2$~pm, \textit{thin
    dashed line}), etalon~2 ($R = 0.95/0.92$, $d = 1.1/0.24$~mm, $\mathrm{FWHM}
    = 2.7/8.9$~pm, \textit{thin dash-dotted line}), and all transmission curves
    combined ($\mathrm{FWHM} = 1.5/4.2$~pm, parasitic-light fraction 0.8/0.4\%,
    \textit{thick solid line}). The transmission profiles are normalized to
    unity at the central wavelength $\lambda_0 = 600.0/400.0$~nm.}
\label{FIG09}
\end{figure}


\subsection{Etalons}

The design of a dual-etalon system requires the consideration of several 
etalon parameters, i.e., absorption, reflectivity, and resulting peak transmission 
and finesse of both etalons, as well as the correct plate-spacing in order 
to obtain the appropriate spectral resolution for a given wavelength range together with  
a low parasitic-light fraction. The latter is defined as the ratio of the intensity 
transmitted out-off band to that of the central transmission peak of the resulting 
Airy-function of the instrument and depends also on the used pre-filter.\cite{2012AN....333..880P} 
The reflectivity of etalons for imaging in the blue spectral region has to be lower to 
increase the FWHM and to accommodate the smaller number of available photons. 
Therefore, the rejection of parasitic light further away from the central wavelength 
$\lambda_0$ becomes more important and narrower interference filters are required.
Interference filters with a FWHM of 0.3--0.8~nm provide a good compromise between parasitic light, 
transmission, and signal-to-noise. Figure~\ref{FIG09} shows the results of such a parameter 
study for BLISS together with the results obtained for the current etalon properties of the GFPI. 
In both panels, the transmission profiles of the individual etalons and the order-sorting 
interference filters ($\rm{FWHM}=0.3$~nm, 40\% transmission) are presented for center wavelengths of 
600~nm (GFPI, upper panel) and 400~nm (BLISS, lower panel), respectively. 
The final BLISS and GFPI transmission profiles (thin solid curve) are multiplied by the transmission 
curve of the pre-filters (thick solid curve) and have a central peak with a FWHM of 1.5~pm and 4.2~pm, respectively. 
The parasitic-light fraction has been calculated according to the wavelength range of the assumed pre-filter, 
i.e., $\pm 3.5$~nm, and amounts to 0.8\% and 0.4\% for GFPI and BLISS, respectively.
The lower parasitic-light fraction of BLISS is a direct result of optimizing the ratio 
of the plate spacings which could be done only partly in case of the GFPI.\cite{2012AN....333..880P}
In principle, a larger parasitic light fraction should be expected in the blue. Our results also show
that only 0.3~nm interference filters will work with BLISS. In contrast to the GFPI etalons with plate 
spacings of $d = 1.1$ and $1.4$~mm ($\mathrm{FWHM} = 2.7$ and $2.1$~pm) and reflectivities of 95\%, 
for BLISS the combination of etalons  with narrower plate spacings of $d = 0.24$ and $0.41$~mm 
($\mathrm{FWHM} = 8.9$ and $5.2$~pm) and lower reflectivities of 92\% will increase the 
light level and result in a theoretical spectral resolution of ${\cal R}\sim
100,000$. A proper wavelength-dependent parameter study like in the case of the GFPI\cite{2012AN....333..880P} requires 
the knowledge of the etalon characteristics provided by the manufacturer, thus the present calculations 
should be extended as soon as this information is available.

\section{Summary and conclusion}

Much effort has been spent to provide with the GFPI a modern first-light post-focus 
instrument for the GREGOR solar telescope. New software and hardware
provide now an easy handling of the instrument with automated procedures for both observation 
and calibration. The GFPI can be controlled with a well-structured
Graphical User Interface (GUI) and underwent an extended comissioning in 2011 and a careful 
science verification throughout 2012. The latter revealed that most of the characteristic 
parameters of the instrument comply with the theoretical expectations. However, we are 
aware of a problem concerning the measured effective spectral resolution of the instrument, 
which warrants further investigation.  

The results from first 2D spectroscopic observations supported by adaptive optics at the end of July 2012
are very promising and suggest that high-cadence 2D spectroscopic observations near 
the diffraction limit are very likely to be obtained in coming observing campaigns at the GREGOR
telescope as soon the apparent stray-light problem is solved. This will also increase significantly
the performance of the adaptive optics system and will finally permit a resolution of about 50~km 
on the solar surface in combination with post-factum image restoration techniques. 
Vector-polarimetric observations will follow in the first half of 2013. 

Possibilities for the extension of polarimetric observations to the entire wavelength range covered 
by the instrument (530--860~nm) are under investigation. We also discussed a future upgrade with new cameras, 
which would both yield higher frame rates and a larger field-of-view. This would be crucial especially 
for the dual-beam spectropolarimetric mode of the GFPI, although the enlargement of the field-of-view would require 
also a re-design of the polarimeter. 

The future combination of the GFPI with its companion, the BLue Imaging Solar 
Spectrometer (BLISS), whose design was presented here in detail, and
the GRIS will permit 
multi-wavelength observations over the entire optical spectral range 380--1600~nm. 
This will allow one to study the connection of different physical processes from 
the lowermost photosphere, i.e., the continuum forming layers, up to the upper 
chromosphere, e.g., the line core of H$\alpha$, at highest spatial
resolution in unprecedented intrinsic detail.

%
%

\acksect*{Acknowledgments}

The 1.5-meter GREGOR solar telescope was build by a German 
consortium under the leadership of the Kiepenheuer-Institut f\"ur Sonnenphysik 
in Freiburg with the Leibniz-Institut f\"ur Astrophysik Potsdam, 
the Institut f\"ur Astrophysik G\"ottingen, and the Max-Planck-Institut f\"ur 
Sonnensystemforschung in Katlenburg-Lindau as partners, and with contributions 
by the Instituto de Astrof\'\i sica de Canarias and the Astronomical Institute 
of the Academy of Sciences of the Czech Republic. CD was supported by
grant DE~787/3-1 of the Deutsche Forschungsgemeinschaft (DFG). We would like to
thank Robert Geissler for his competent and extended help during the science
verification.

%
%


\vfill

\newpage

\begin{biography}[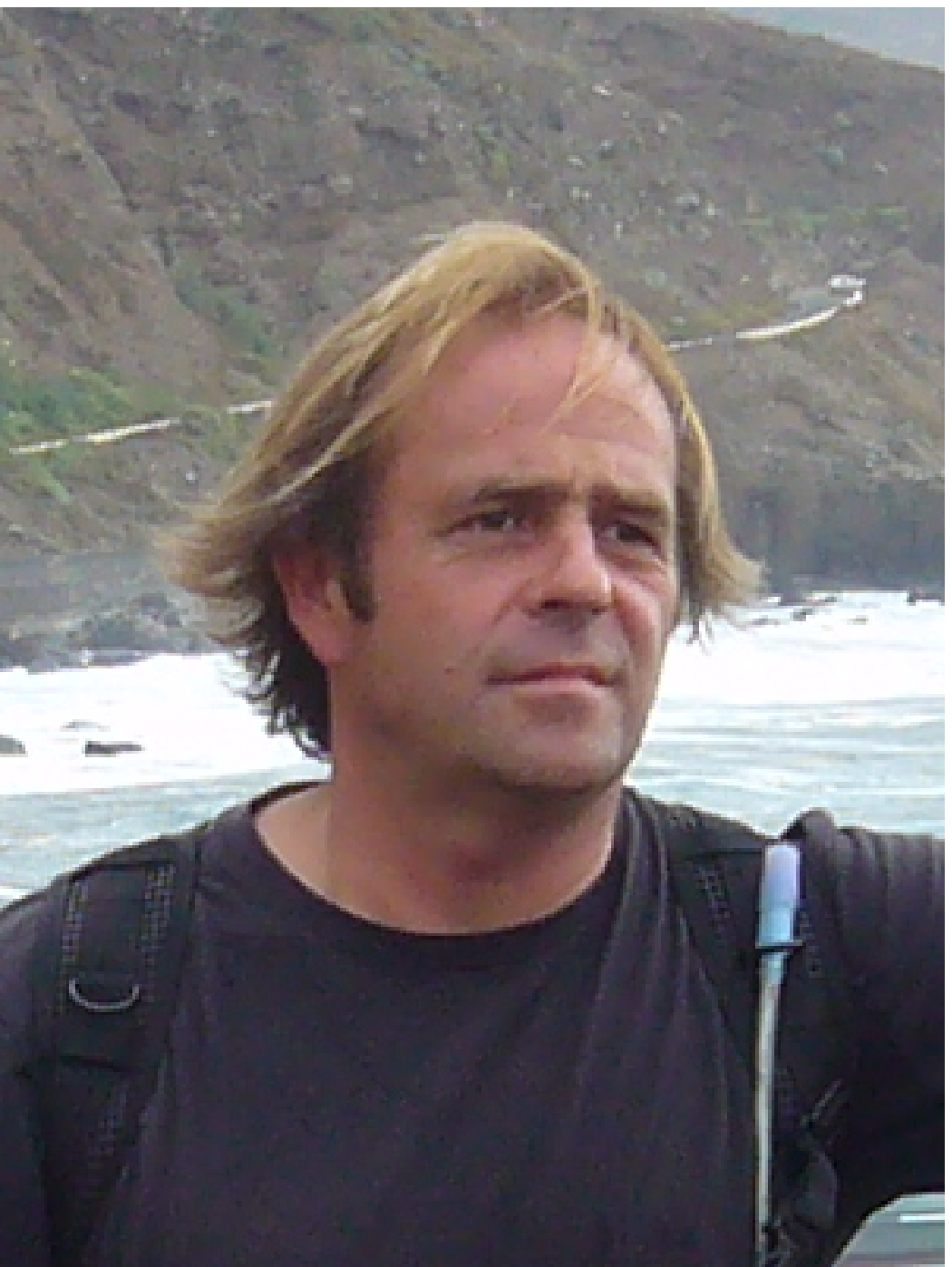]{Klaus G. Puschmann}
performed his thesis at the Instituto de Astrof{\'i}sica de Canarias (IAC) and
received his doctoral degree from the Karl-Franzens Universit\"at Graz in 2002.
After post-doctoral positions at the Institut f\"ur Astrophysik G\"ottingen and the IAC,
he is presently employed at the Leibniz-Institut f\"ur Astrophysik Potsdam. His research 
interests include image reconstruction, spectropolarimetry and Stokes inversion of sunspot spectra.
He is member of the GREGOR project with focus on the development of the GREGOR Fabry-P\'erot 
Interferometer.
\end{biography}
\begin{biography}[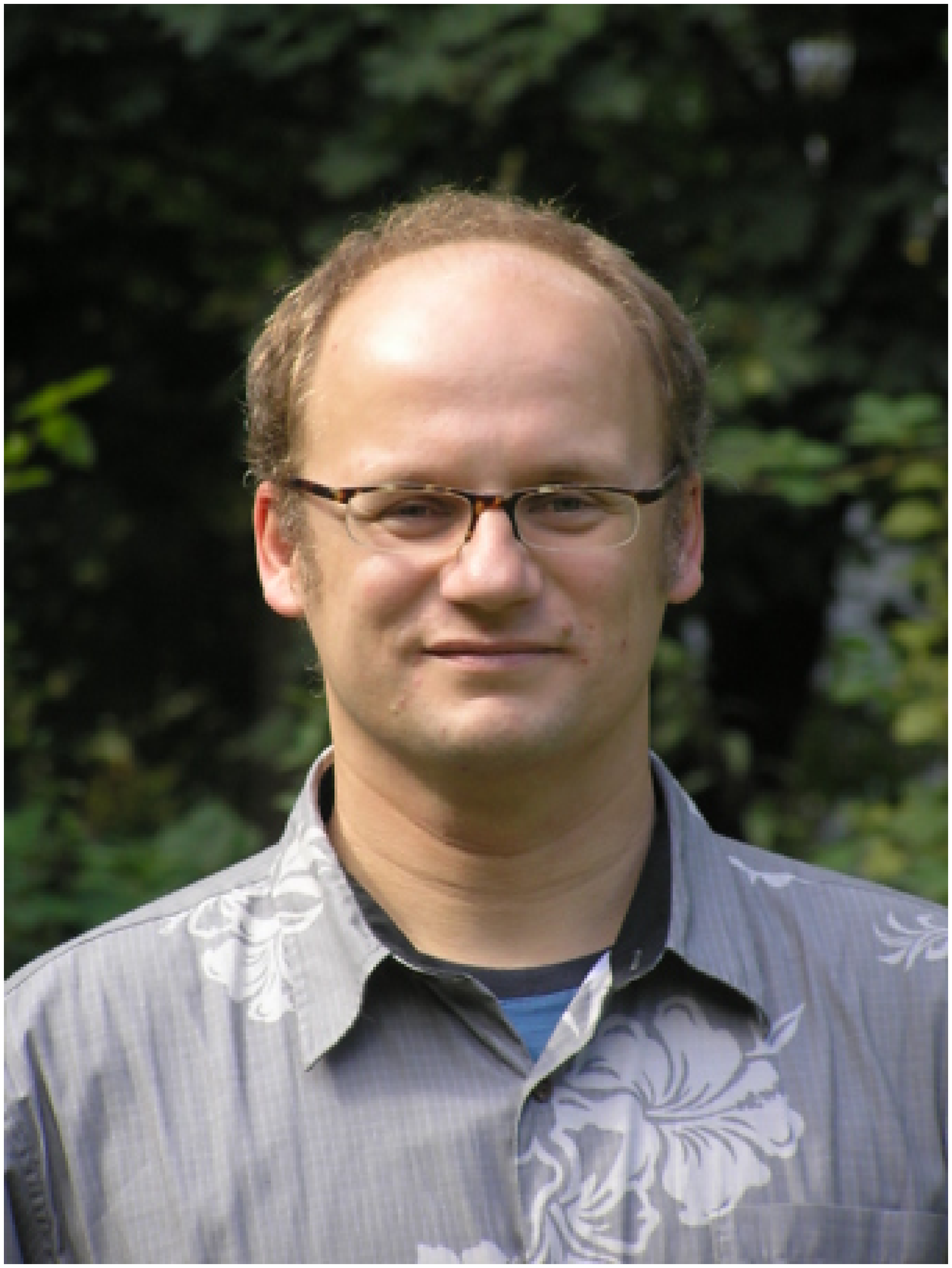]{Carsten Denker}
received his doctoral degree from the Georg-August-Universit\"at G\"ottingen in
1996. At the Leibniz-Institut f\"ur Astrophysik Potsdam, he is the head of the
Optical Solar Physics Group and the Solar Observatory Einstein Tower. He is
teaching astronomy, astrophysics, and solar physics at the Universit\"at Potsdam
and the Humboldt-Universit\"at zu Berlin. His main research interests are in the
areas of solar physics, instrumentation for high-resolution solar observations,
and image restoration.
\end{biography}
\begin{biography}[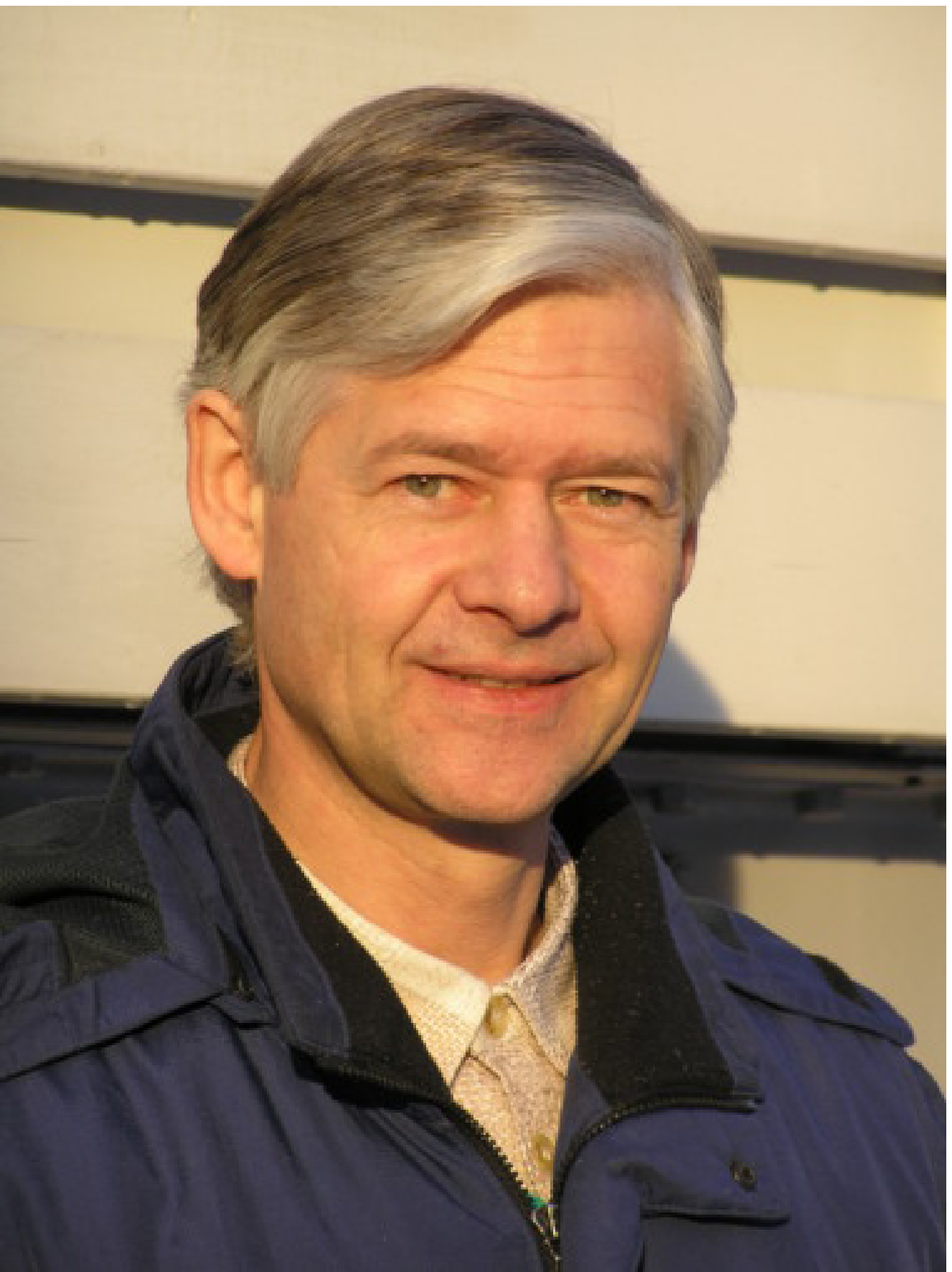]{Horst Balthasar}
studied physics at the Georg-August-Universit\"at G\"ottingen
where he got his doctoral degree in 1984. After scientific stations
at the Instituto de Astrof{\'\i}sica de Canarias, the Universit\"at G\"ottingen,
the Fachhochschule Wiesbaden and the Kiepenheuer Institut f\"ur Sonnenphysik 
Freiburg, he works since 1997 as a scientist at the Leibniz-Institut f\"ur 
Astrophysik Potsdam. His main scientific interest is the investigation of the
dynamics and magnetic field in sunspots.
\end{biography}
\begin{biography}[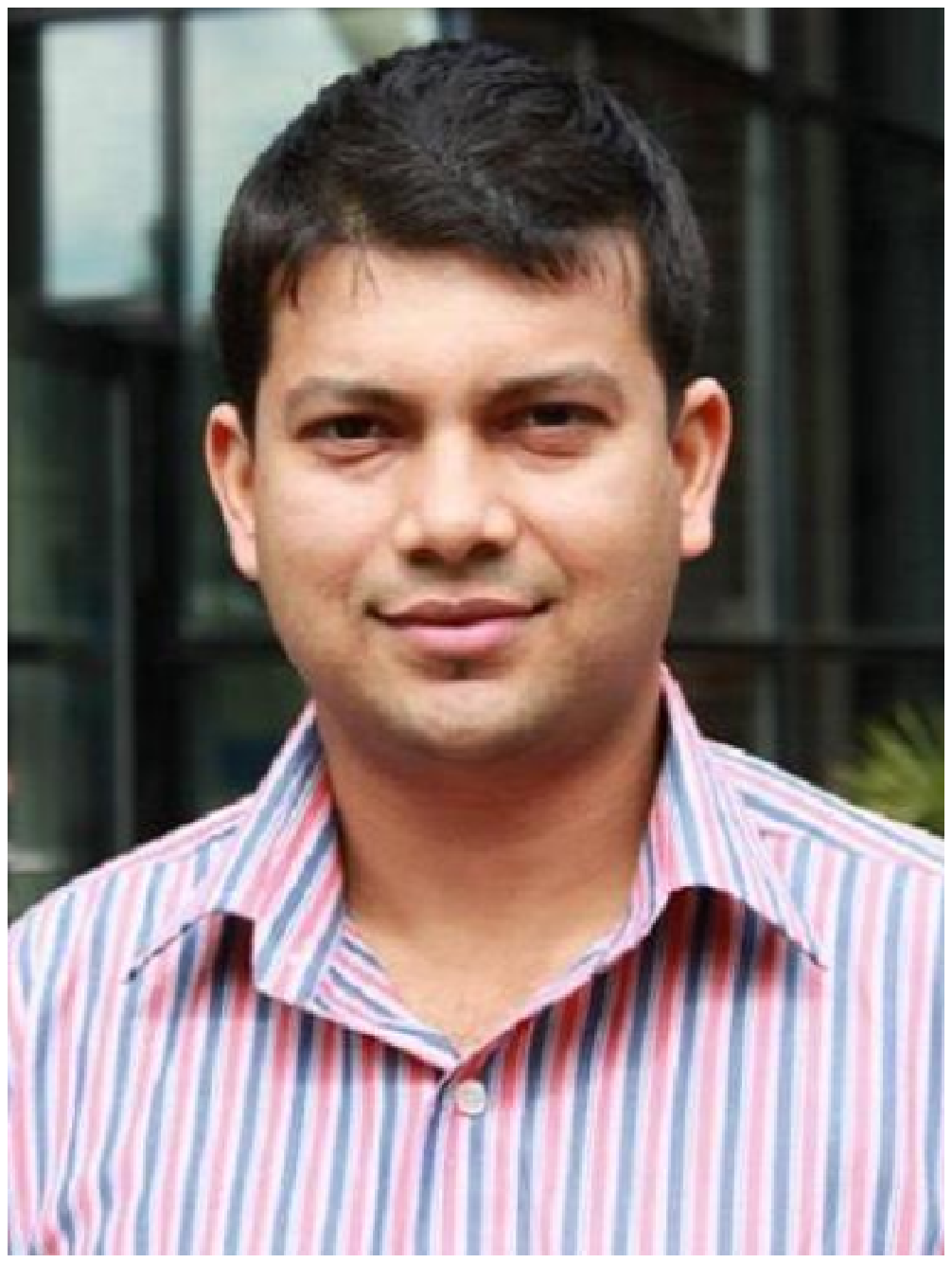]{Rohan E. Louis} 
completed his PhD at the Udaipur Solar Observatory, India in 2010. He 
is currently working as a post-doctoral fellow at the Leibniz-Institut 
f\"ur Astrophysik Potsdam where he is involved in scientific investigations 
of high-resolution imaging-spectropolarimetric observations from the GREGOR 
Fabry-P\'erot Interferometer. His research interests include sunspot fine 
structure, photospheric magnetic fields, solar adaptive optics, and Stokes 
inversion codes.
\end{biography}
\begin{biography}[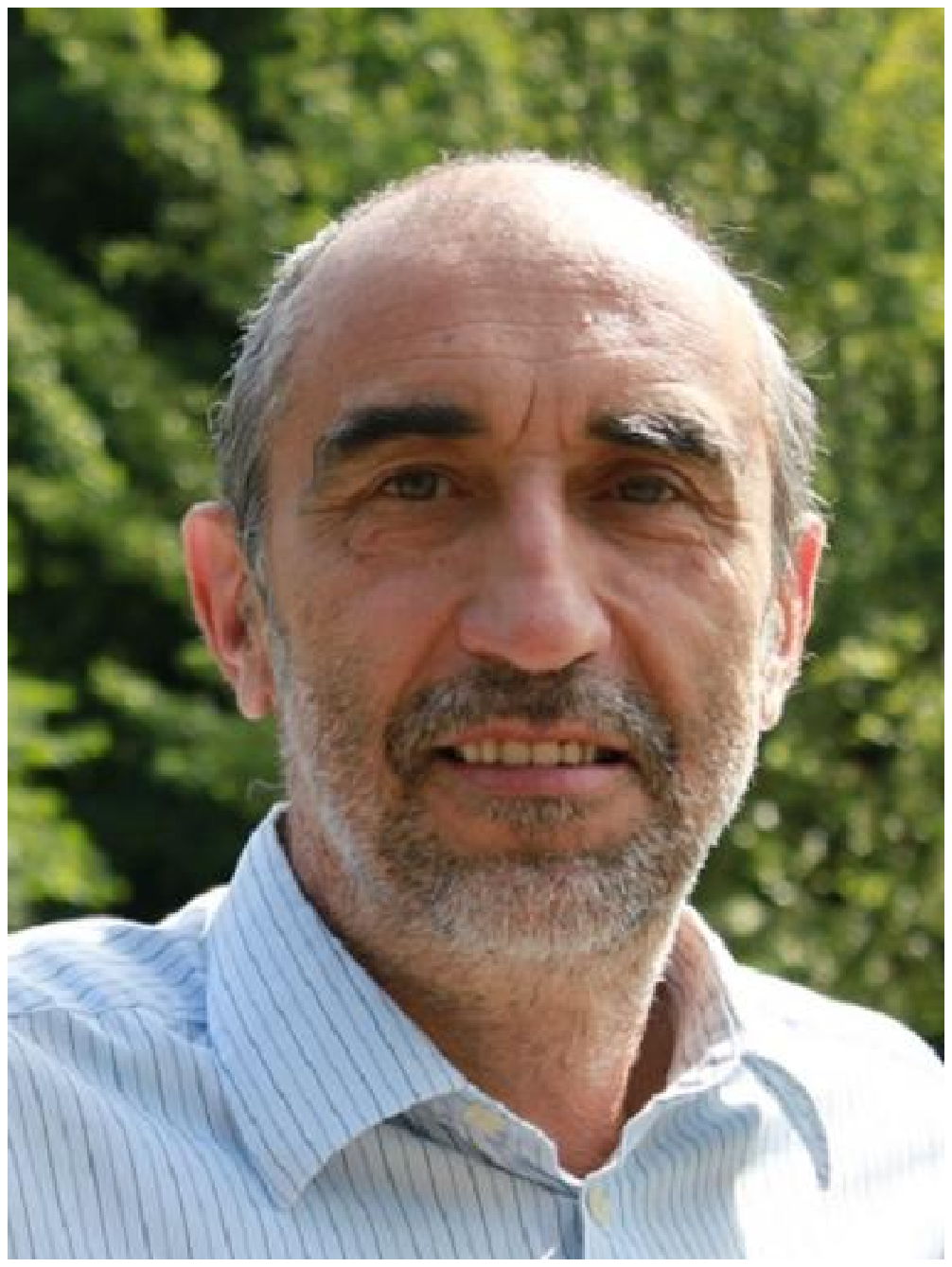]{Emil Popow}
studied physics at the Friedrich-Schiller-Universit\"at Jena and the Humboldt-Universit\"at zu Berlin.
He is employee of the Leibniz-Institut f\"ur Astrophysik Potsdam, where he worked on the development 
of scientific instruments and astronomical instrumentation. In 2003, he became the head of AIP's technical division.
\end{biography}
\begin{biography}[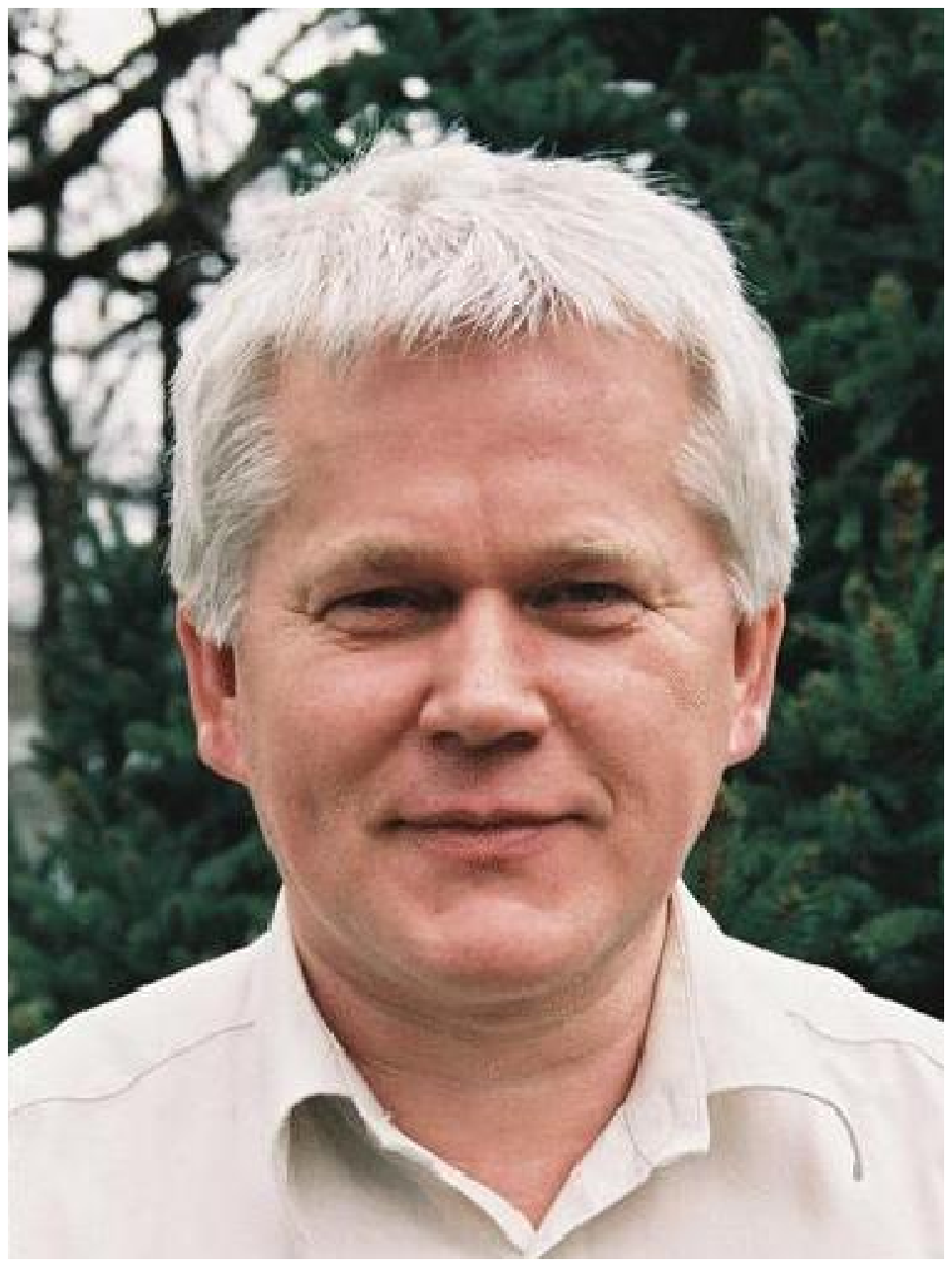]{Manfred Woche}
obtained his diploma in physics at the Friedrich-Schiller-Universit\"at Jena in 1976.
Until 1996 he was employed at the Karl-Schwarzschild-Observatorium Tautenburg with core responsibilities in 
spectroscopy and instrumentation.
>From 1996 until 2001 he worked in optical design and astronomical instrumentation for different institutes, 
especially for the Max-Planck-Institut f\"ur extraterrestrische Physik Garching and the Skinakas Observatory Crete. 
Since 2001 he is responsible for the optical design of astronomical instrumentation at the Leibniz-Institut f\"ur 
Astrophysik Potsdam.
\end{biography}
\begin{biography}[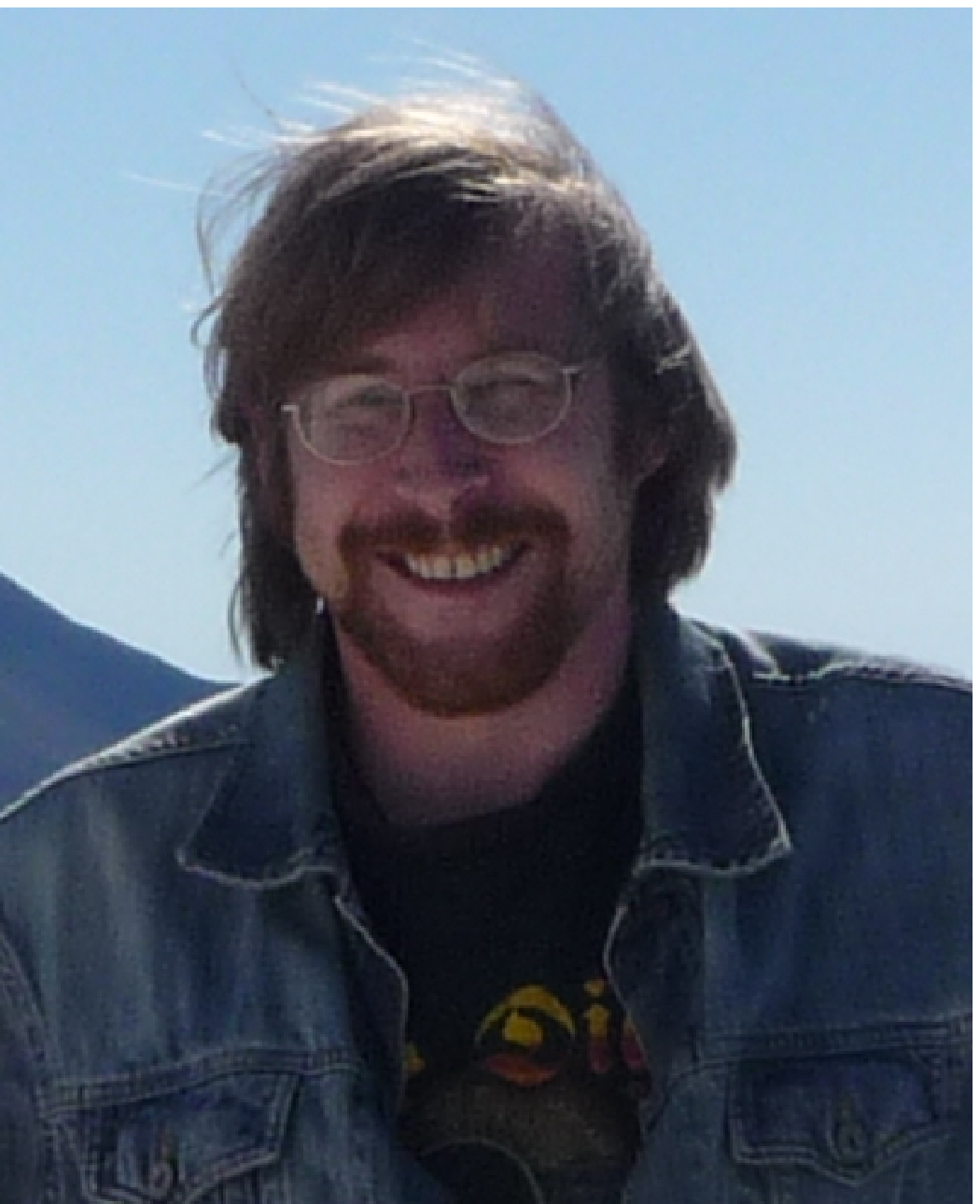]{Christian Beck}
obtained his PhD at the Albert-Ludwigs-Universit{\"a}t Freiburg in 2006. 
He worked as a post-doctoral researcher at the Instituto de Astrof{\'i}sica de 
Canarias from 2006 to 2012 with a research focus on instrumentation, the structure 
of sunspots, and the solar chromosphere. He recently joined the National Solar 
Observatory in Sunspot, New Mexico, as an instrument scientist.
\end{biography}
\begin{biography}[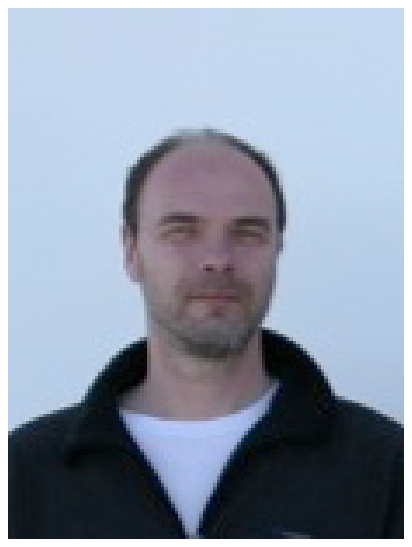]{Thomas Seelemann}
studied physics at the Georg-August-Universit\"at G\"ottingen. For the diploma thesis and doctoral dissertation 
he worked at the Max-Planck Institut f\"ur Str\"omungsforschung in G\"ottingen in the field of molecular physics and 
received his doctoral degree in 1980. Since then, he is employed at LaVision GmbH, where he works on cameras and their integration into 
dedicated experimental applications.
\end{biography}
\begin{biography}[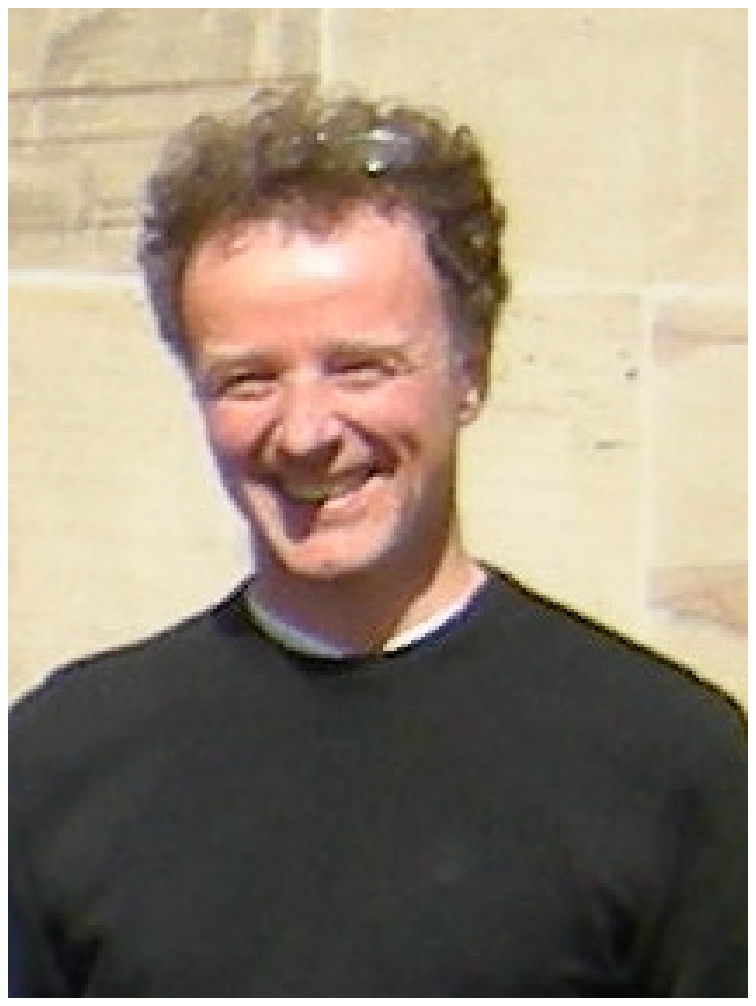]{Reiner Volkmer}
received his doctoral degree in physics from the Georg-August-Universit\"at G\"ottingen in 1995.
After his PhD, he was project manager of the data handling system for IBIS on board of the INTEGRAL satellite
at the Universit\"at T\"ubingen. Later on, he was manager of the GREGOR project and is now leading the development 
of the image stabilization system for the PHI instrument on board the Solar Orbiter. 
\end{biography}
\end{document}